\documentclass[10pt,letter]{IEEEtran}

\usepackage{times}
\usepackage{graphicx}
\usepackage{url}
\usepackage{epsfig}
\usepackage{subcaption}

\usepackage{algorithm}
\usepackage{algorithmicx}
\usepackage[noend]{algpseudocode}
\usepackage{booktabs}
\usepackage{xspace}
\usepackage{multirow}

\newcommand\ignore[1]{}
\newcommand\emc{\emph{AC}\xspace}
\newcommand\cc{C\&C}
\newcommand\detectcc{{\sf Detect\_C\&C}}
\newcommand\computescore{{\sf Compute\_SimScore}}
\newcommand\nohosts{{\small \sf NoHosts}}
\newcommand\noautohosts{{\small \sf AutoHosts}}
\newcommand\noreferrer{{\small \sf NoRef}}
\newcommand\rareua{{\small \sf RareUA}}
\newcommand\domainage{{\small \sf DomAge}}
\newcommand\domainval{{\small \sf DomValidity}}
\newcommand\iptwofour{{\small \sf IP24}}
\newcommand\ipsixteen{{\small \sf IP16}}
\newcommand\dominterval{{\small \sf DomInterval}}
\newcommand\tscore{{T_{s}}}

\newcommand{\secref}[1]{\S\ref{#1}}

\title{Detection of Early-Stage Enterprise Infection by Mining Large-Scale Log Data}

\author{
\IEEEauthorblockN{Alina Oprea, Zhou Li,}
\IEEEauthorblockA{RSA Laboratories, Cambridge, MA, USA}
\\
\and
\IEEEauthorblockN{Ting-Fang Yen,}
\IEEEauthorblockA{E8 Security, Palo Alto, CA, USA}
\and
\\
\IEEEauthorblockN{Sang Chin,}
\IEEEauthorblockA{Draper Laboratory, Cambridge, MA, USA}
\and
\\
\IEEEauthorblockN{Sumayah Alrwais,}
\IEEEauthorblockA{Indiana University, Bloomington, IN, USA}
}

\ignore{
\makeatletter
\renewcommand{\@maketitle}{
\newpage
 \null
 \vskip 2em%
 \begin{center}%
  {\LARGE \ttlfnt \@title \par}%
 \end{center}%
\vskip 3.5em%      %%%%% Adjust this parameter to control space between
 \par} \makeatother
}

\begin{document}

\maketitle
\pagenumbering{arabic}

\begin{abstract}
Recent years have seen the rise of more sophisticated attacks including advanced persistent threats (APTs)~\cite{Stuxnet,RSA,NYTimes,Target} which pose severe risks to organizations and governments by targeting confidential proprietary information. Additionally, new malware strains are appearing at a higher rate than ever before~\cite{Panda}. Since many of these malware are designed to evade existing security products, traditional defenses deployed by most enterprises today, e.g., anti-virus, firewalls, intrusion detection systems, often fail at detecting infections at an early stage.

We address the problem of  detecting early-stage infection in an enterprise setting by proposing a new framework based on belief propagation inspired from graph theory. Belief propagation can be used either with ``seeds'' of compromised hosts or malicious domains (provided by the enterprise security operation center -- SOC) or without any seeds. In the latter case we develop a detector of \cc\ communication particularly tailored to enterprises which can detect a stealthy compromise of only a single host communicating with the \cc\ server.

We demonstrate that our techniques perform well on detecting enterprise infections.  We achieve high accuracy with low false detection and false negative rates on two months of anonymized DNS logs released by Los Alamos National Lab (LANL), which include APT infection attacks simulated by LANL domain experts. We also apply our algorithms to 38TB of real-world web proxy logs collected at the border of a large enterprise. Through careful manual investigation in collaboration with the enterprise SOC, we show that our techniques identified hundreds of malicious domains overlooked by state-of-the-art security products.

%Our techniques identified hundreds of suspicious domains overlooked by state-of-the-art security products. Through careful manual investigation in collaboration with the enterprise SOC, we confirm that a large fraction of detected domains are used for malicious purposes.

%We show the effectiveness of our techniques by identifying hundreds of suspicious domains contacted by enterprise hosts overlooked by state-of-the-art security products.

\ignore{
APT-like infection attacks simulated by LANL domain experts. We achieve high accuracy with low false discoveryand false negative rates at identifying the LANL simulated attacks among two month of anonymized DNS logs released to the community. 
}

\ignore{
We leverage an anonymized dataset publicly released by Los Alamos National Lab (LANL) that includes two months of  DNS logs collected at LANL with 20 simulated APT attacks overlaid onto it. We propose a new graph-theoretic framework based on belief propagation to detect the attacks and demonstrate that our detector has high accuracy while the incurred false positive and false negative rates are low. We then apply our techniques to real-world web proxy logs collected at the border of a large enterprise network. We show the effectiveness of our techniques by identifying hundreds of suspicious domains contacted by enterprise hosts overlooked by state-of-the-art security products. Through careful manual investigation, we confirm that a large fraction of detected domains are used for malicious purposes.
}

\end{abstract}

%\keywords{Advanced Persistent Threats, Data Analysis, Belief Propagation}

\section{Introduction}

The cybersecurity landscape is evolving constantly. More sophisticated attacks including Advanced Persistent Threats (APTs)~\cite{Stuxnet,RSA,NYTimes,Target} have emerged recently targeting organizations' intellectual property, financial assets, and national security information. Well-funded attackers use advanced tools and manually orchestrate their campaigns to adapt to the victim's environment and maintain low profiles of activity.
Additionally there are also more malware than ever before. A whitepaper published by Panda Labs~\cite{Panda} found 30 million new malware strains in circulation in 2013 alone, at an average of 82,000 malware a day.
Many of these are variants of known malware designed to evade existing security products, such that existing defenses, e.g., anti-virus, firewalls, intrusion detection systems, often fail at detecting infections at an early stage.

However, certain \emph{infection patterns} still persist across malware variants and families due to the typical infection vectors used by attackers. For example, during the malware \emph{delivery stage}, victim hosts often visit several domains under the attacker's control within a short period of time as a result of redirection techniques employed by attackers to protect their malicious infrastructures~\cite{ShadyPath}. After delivery, backdoors are installed on the compromised machines to allow \emph{footholds} into the targeted organization~\cite{APT1}, where the machines initiate outbound connections regularly to a command-and-control server to receive instructions from the attacker. Malware communications commonly take place over HTTP/HTTPS, since web traffic is typically allowed by firewalls. More importantly, domains used in the same attack campaign are often related, meaning that they may share locality in either IP address space, time of access or set of hosts contacting them. These patterns of infections have been observed in advanced targeted attacks (e.g., APT1 group~\cite{APT1}, Shady RAT~\cite{Shady}, Mirage~\cite{Mirage}), as well as botnet infections (e.g., Zeus, Citadel~\cite{Botnets} and ZeroAccess~\cite{ZeroAccess}).

In this work, we leverage these observations to detect early-stage malware infections in enterprise networks. Our focus on enterprises stems from their unique point of view and new challenges present in those networks. As the victims of many cyber attacks, enterprises are pressed on detecting infections early to prevent further damage. \ignore{, e.g., the exfiltration of sensitive data, which may be costly or even impossible to remediate. }However, the amount of network traffic generated by a large enterprise \ignore{(e.g., on the order of 10,000 to 100,000 employees)} can be terabytes per day, requiring extremely efficient analysis methods to maintain a reasonable detection time. Yet, even though there is much data, it is of limited scope --- only containing traffic as observed from that enterprise, making prior approaches using ISP-level data (e.g.,~\cite{Bilge2011,Antonakakis2011}) inapplicable.

%since such information is sensitive and rarely shared with other parties. Prior works on detecting malicious activities with a wide set of vantage points, such as ISP-level DNS records~\cite{}, are hence not applicable in this setting.

\ignore{
Detection and response, as performed in enterprises today, are largely the responsibility of the enterprise Security Operations Center (SOC). The SOC consists of a team of security analysts that monitors the network activities inside the enterprise network, often relying on commercial blacklists or external intelligence sources (i.e., Indicators of Compromise, or IOC) to identify malicious domains and infected hosts, followed by manual investigation to determine the scope of the threat. As IOCs are by no means complete, the investigation phase is particularly labor-intensive. It is this process that we aim to facilitate in this paper.
}

We propose a graph-theoretic framework based on belief propagation~\cite{Pearl82} to identify \emph{small communities} of related domains that are indicative of early-stage malware infections. We first restrict our attention to traffic destined to \emph{rare destinations}. These are ``new'' domains, not visited before by any host in the organization within an observation window (and thus more likely to be associated with suspicious activity), and contacted by a small number of internal hosts (since we expect the initial infection to be small). In each iteration of our belief propagation algorithm, the rare domains are scored according to several features and similarity with domains detected in previous iterations.

%the degree to which it exhibits command-and-control (/cc/) behaviors, or its similarity to other identified suspicious domains.

Our algorithm can be applied either with ``hints'' (starting from ``seeds'' of known compromised hosts or domains), or without (when no information about compromised hosts or domains is given). In the first case, seeds can be obtained from commercial blacklists and external intelligence sources containing Indicators of Compromise (IOCs) that the enterprise security operations center (SOC) has access to. Currently, SOC security analysts  manually investigate incidents starting from IOCs, and we aim here to facilitate this process. In the latter case, our method first identifies automated, regular \cc\ communications that can then be used as seeds. Our \cc\ detector leverages unique properties of enterprise networks (e.g., popularity of user-agent strings and web referer information in the HTTP traffic) as well as features utilized in previous works (e.g., timing patterns, domain age and registration validity) and it can detect  a single compromised host.

We demonstrate the effectiveness of our techniques on two different datasets, one containing DNS records and the other web proxy logs. The first consists of two months of anonymized DNS records from Los Alamos National Lab (LANL) in early 2013 amounting to 1.15TB. \ignore{There are a total of 3.81 billion DNS queries and 3.89 billion DNS responses, amounting to 1.15 TB. In addition to real-world DNS traffic, }This dataset also includes 20 independent APT-like infection attacks simulated by LANL domain experts and was released along with a challenge problem to the community (\emph{APT Infection Discovery using DNS Data~\cite{LANL}}) requesting methods to detect compromised internal hosts and external domains in the simulated attacks. The challenge included ``hints'' of varying details (e.g., one or multiple known compromised hosts), as well as answers for validation. Our techniques proved effective at detecting the LANL simulated attacks achieving an overall 98.33\% true detection rate, at the cost of 1.67\% false detection rate and 6.25\% false negative rate.

Our second dataset contains two months of web proxy logs collected from a large enterprise in early 2014. Detecting malicious infections in this dataset proved to be more challenging due to its large scale (38TB of data), inconsistent information (host IP addresses are dynamically assigned using DHCP in most cases), and also the lack of ground truth. Through careful manual analysis in collaboration with the enterprise SOC, we identify hundreds of malicious domains not previously detected by state-of-the-art security products deployed on the enterprise network. Interestingly, a large number of these (98 distinct domains) are entirely new discoveries, not reported yet by VirusTotal several months after we detected them.  This demonstrates the ability of our techniques to detect entirely new, previously unknown attacks.

To summarize our main contributions in the paper are:

\vspace{2pt}

\noindent\textbf{Belief propagation framework for detecting enterprise infection.} We develop a graph-theoretic framework based on belief propagation for detection of early-stage enterprise infections. Given ``seed'' hosts or domains, we automatically infer other compromised hosts and related malicious domains likely part of the same campaign. Our approach uniquely leverages relationships among domains contacted in multiple stages of the infection process. 

\vspace{2pt}

\noindent\textbf{Detector of \cc\ communication in enterprise.} By exploiting novel enterprise-specific features and combining them with features used in previous work, we build a detector of \cc\ communication tailored to an enterprise setting. Domains labeled as potential \cc\ can be seeded in the belief propagation algorithm to detect other related domains.

\vspace{2pt}

\noindent\textbf{Solve the LANL challenge.} We apply the belief propagation algorithm to the LANL challenge and identify the malicious domains in the 20 simulated campaigns with high accuracy and low false detection and false negative rates.

\vspace{2pt}

\noindent\textbf{Evaluate on real-world data from large enterprise.} We apply our solution to a large dataset (38.41 TB) of web proxy logs  collected at an enterprise's network border. We identify hundreds of suspicious domains contacted by internal enterprise hosts which were not detected previously by state-of-the-art security products. Among the 375 domains detected in total over a month, 289 (accounting for 77.07\%) are confirmed malicious or suspicious through careful manual investigation. While 191 (50.93\%) are also reported by VirusTotal (but unknown to the enterprise of our study), we identify 98 (26.13\%) that are entirely new discoveries (not reported by VirusTotal or the enterprise).

%A large fraction of these are confirmed malicious through careful manual investigation.

%\vspace{2pt}
%
%\noindent\textbf{New findings of confirmed malicious activities.} We identify hundreds of malicious domains contacted by internal enterprise hosts which were not detected previously by state-of-the-art security products.

\section{Problem statement}
\label{sec:statement}

Our goal is to detect malware infection within an organization in early stages of the campaign. \ignore{Enterprise networks are difficult to penetrate as different security products are deployed for protecting the infrastructure. However, more sophisticated attacks (APTs) are specifically crafted to adapt to the targeted environment, maintain stealthy operation to avoid detection and persist for months or even years in the victim organization to reach their objectives.} We describe below the characteristics of common enterprise infections (due to either targeted or opportunistic attacks), why existing solutions fail against such threats and the challenges we had to overcome for detecting them.

\subsection{Enterprise Infections}

Common infection vectors for targeted attacks are social engineering~\cite{APT1} and compromise of legitimate sites~\cite{WebSense}. In the case of social engineering, attackers craft legitimate-looking spear-phishing email addressed to several employees within the targeted organization including a malicious attachment or a hyperlink to a malicious file. Attack vectors employed by mainstream malware include spam emails, USB drives, and a variety of web-based attacks (e.g., drive by download, clickjacking, malvertising, etc.). Many of these attacks (both targeted and mainstream) have a common pattern during early-stage infection~\cite{APT1,Shady,Mirage,Botnets}:

\ignore{
Another infection vector gaining popularity recently is ``watering-hole'' attacks~\cite{VoHo}, in which attackers compromise legitimate sites likely visited by employees of the targeted organization (to either serve malicious payload directly or redirect to a malicious site).
 }

\vspace{2pt}

\noindent {\bf Delivery stage:} During delivery, the victim machine gets the malicious payload, for example by an email attachment, or drive-by-download attack, etc. Many times, the first-stage malware is generic and needs to download additional malware (second stage) specifically crafted for the victim environment~\cite{Thonnard12}.

\vspace{2pt}

\noindent {\bf Establishing foothold:} After delivery a backdoor is usually installed on the victim's machine and the attacker establishes a foothold within the organization~\cite{APT1}. In almost all cases, backdoors initiate outbound connections to evade firewalls that block connections from outside the network. Most communications go through HTTP or HTTPs since these ports are allowed by most enterprise firewalls~\cite{RSAkillchain, ExecScent}.

\vspace{2pt}

\noindent {\bf Command-and-control (\cc):} Typically, backdoors connect regularly to the command-and-control center operated by attackers to receive further instructions and allow attackers backdoor access into the victim environment~\cite{APT1,Botnets}.

Based on these infection patterns, we extract several common characteristics of enterprise infections:

\vspace{2pt}

\noindent {\bf Small scale:} Attackers are motivated to maintain stealthy operations to avoid detection by security products deployed within enterprise perimeter and we focus on detecting infections with small scale. In some cases a single host might get infected and communicate with the \cc\ center.

%Comparing to the enormous traces generated from the enterprise's normal operations, the volume of traffic generated by compromised hosts is extremely small.

\vspace{2pt}

\noindent {\bf Uncommon domains:} Attackers tend to use uncommon destinations for different stages of the campaign (e.g., delivery, \cc). To examine the popularity of domains used in attacks, we obtained a list of 14,915 IOCs reported between 2011 and 2014 from the SOC of a large enterprise. None of the indicators are among the Alexa top one million most popular domains~\cite{Alexa}. Additionally, \cite{APT1} points out that attackers use more frequently domain names rather than direct IP connections for their \cc\ communication so that they can dynamically flux the domains. Among the 14,915 IOCs in the enterprise list, the vast majority (13,232 or 88.71\%) are domain names.

\ignore{
While attackers could use either external IP addresses or domains to represent \cc servers, they are more likely to use domains. As pointed out in~\cite{APT1}, ``by using Fully-Qualified Domain Names (FQDNs) rather than hardcoded IP addresses as \cc\ addresses, attackers may dynamically decide where to direct \cc\ connections from a given backdoor''.

To examine the popularity of domains used in attacks, we obtained a list of IOCs reported between 2011 and 2014 from a security company, including 14,915 high-risk indicators. Among them, 13,232 (88.71\%) are domain names while 1,239 (8.31\%) are IP addresses. Besides, attackers prefer to register and use uncommon domains in this stage instead of abusing common ones (like \texttt{google.com}) in order to have better control. Again, we examine the same list of IoC and found none of the indicator domains are popular domains among Alexa top 1M domain list~\cite{Alexa}.
}

\vspace{2pt}

\noindent {\bf HTTP/HTTPs Communication.} The communications between malware and \cc\ servers is typically done through HTTP or HTTPs since other ports are blocked by enterprise firewalls~\cite{RSAkillchain, ExecScent}.

\vspace{2pt}

\noindent {\bf Communities of domains:} A compromised host usually contacts several malicious domains within a relatively short time interval. For instance, a user clicking on an embedded link in an email might visit the front-end attacker site, get redirected to a site hosting malicious payload and shortly after the backdoor is established will initiate the first connection to the \cc\ server. These domains form \emph{small communities} exhibiting similarity in connection timing, set of hosts contacting them (if multiple hosts are infected in the same campaign) and sometimes proximity in IP address space~\cite{HaoIMC11, APT1}.

%in the bipartite graph including connections between internal hosts and external destinations.

\vspace{2pt}

\noindent {\bf Automated \cc\ communication:} Backdoors typically communicate with \cc\ servers on a regular basis to allow attackers access into the victim environment. In many publicized APT campaigns (e.g., NightDragon~\cite{CommandFive}, Mirage~\cite{Mirage}, Shady RAT~\cite{Shady}) as well as botnet infections (e.g., Zeus, Citadel~\cite{Botnets}, ZeroAccess~\cite{ZeroAccess}), \cc\ communication occurs at regular time intervals (minutes or hours). We also examined malware samples provided by Mandiant on the APT1 group to test their communication patterns. Among 43 backdoor samples, only 4 exhibit randomized communication patterns while the remaining ones periodically communicate back to \cc\ servers (with small variation between connections).

\ignore{
Compromised hosts usually communicate to the \cc\ servers \textbf{regularly}. It is reported that the malware used by known APT campaigns, like NightDragon~\cite{CommandFive}, Mirage~\cite{Mirage} and advanced RAT (Remote Access Trojan) like IEXPL0RE RAT~\cite{IEXPL0RE} communicates to \cc\ servers at regular time interval. Also reported by Mandiant, one notorious cyber-criminal group, APT1, plants backdoors on victims' machines which ``regularly connect out to hop points''~\cite{APT1}. We examined malware samples provided by Mandiant on APT1 to confirm this. Among the 43 malwares used in infection stage, only 4 exhibit randomized communication pattern while the remaining ones periodically communicate back to \cc\ servers (with small amount of randomization between connections).
}

\subsection{Current Defenses}

Large enterprises deploy different security products (anti-virus, intrusion-detection, firewalls, etc.). Detection and response are largely the responsibility of the Security Operations Center (SOC). The SOC consists of a team of security analysts that monitors the network activities inside the enterprise network, often relying on commercial blacklists or external intelligence sources (i.e., IOCs) to identify malicious domains and infected hosts, followed by manual investigation to determine the scope of the threat. As IOCs are by no means complete, the investigation phase is particularly labor-intensive. It is this process that we aim to facilitate in this paper.

\ignore{Large enterprises deploy many different security products (intrusion-detection, firewalls, etc.) and have access to external intelligence sources containing commercial blacklists and IOCs. These are lists of domains and IP addresses known to be associated with malicious activity. Blacklists and IOCs are by no means complete, but they can be used as starting points for investigation. Many times, the enterprise SOC receives alerts about internal hosts contacting a blacklisted or IOC destination. Currently the investigation process by SOC is done manually by human analysts leveraging various log data collected in the enterprise, and additional external resources. We aim at automating this process with our belief propagation framework.
}

Systems to detect botnet communication or malicious domains have been extensively proposed in the literature. Some of them (e.g., \cite{BotFinder, JACKSTRAWS}) require malware samples for training. Unsupervised systems (e.g., \cite{BotMiner, BotSniffer, Yen08}) typically require multiple synchronized hosts compromised by the same malware, and do not scale to large networks. ExecScent~\cite{ExecScent} identifies \cc\ domains from enterprise web traffic, though it requires malware samples as input to generate traffic templates. Our approach is also tailored to large enterprise networks like ExecScent, but does not require malware binaries, and can identify multiple related malicious domains used in the campaign (not only the \cc\ stage).

\ignore{
Our detection algorithms (like ExecScent~\cite{ExecScent}) are particularly tailored to a large enterprise network and can scale to processing terabytes of log data. Our approach is orthogonal though to ExecScent (which requires malware samples as input) in that it works either with an IOC provided as input by SOC or without any seeds. It thus has the ability of detecting \emph{new, unknown} campaigns that have certain common infection patterns.
}

\subsection{Challenges}

There were several challenges we had to overcome in the process of developing our detection methodology. First, security products deployed within the enterprise perimeter record large volumes of data daily. For example, the two datasets we used to evaluate our system are 1.15 TB and 38.14 TB, respectively. To perform efficient analysis, we describe in~\secref{sec:datareduction} a suite of techniques that reduce the data volume by an order of magnitude while retaining the valuable information about communication of internal hosts to external domains.

%In practical situations we have to deal with incomplete infection data as machines might get infected outside the corporate network and thus we don't always have visibility in all infection stages.

In practical situations, we have to deal with different log formats generated by a variety of deployed security products. We focus on general patterns of infections that is common in various types of network data (e.g., NetFlow, DNS logs, web proxies logs, full packet capture) and demonstrate the effectiveness of the algorithms on two distinct datasets (DNS logs in~\secref{sec:lanl} and web proxies logs in~\secref{sec:eval_emc}). In the web proxies log dataset we enrich the set of features with context information available in HTTP connections.

Finally, while most existing detection systems focus on general malware, we aim to also tackle sophisticated infections which could be part of a targeted attack. APT attacks are extremely stealthy, maintain low profiles and adapt to the victim environment, leaving only small amount of evidence in the log data. We develop an algorithm for identifying suspicious \cc\ domains in~\secref{sec:cc} (even when contacted by a single host) requiring no prior knowledge of the malware sample. We use these domains as seeds to identify other related malicious domains and compromised hosts part of the same campaign through belief propagation. In the case when IOCs are provided (which is common in enterprise settings), they can be used as seeds in the same belief propagation algorithm. Our algorithms are unique in identifying relationships among domains used in different infection stages, e.g., delivery, \cc. The evaluation on LANL dataset shows that our approach is able to detect simulated APT infections and thus has the potential of detecting targeted attacks in the wild.

%To the best of our knowledge, most existing approaches designed to counteract general malicious activities are unsuccessful when faced with sophisticated attacks like targeted attacks. We describe below the typical operation of APT campaigns, the adversarial model considered in the paper and the challenges we had to overcome in building our detection techniques. 

\section{Methodology}
\label{sec:methodology}

In this section, we provide an overview of our approach to detecting early-stage enterprise infection. Our system analyzes log data collected at the enterprise border on a regular basis (e.g., daily), maintains profiles of normal activity within the enterprise, and detects malware infections by exploiting the relationship between suspicious external destinations used in different infection stages. We first describe the datasets used in this study, then introduce our main framework based on belief propagation, the features for detecting suspicious external destinations, and conclude with an overview of the system operation. Details on our techniques and features are given in the next section.

%by analyzing large-scale log data collected at the border of an enterprise.

\subsection{Datasets}

% TF: This paragraph should come earlier in the paper... probably where we state our goals?
%Our main goal is detecting suspicious external destinations contacted by internal hosts in an organization. At the high level our system analyzes log data collected at the enterprise border on a regular basis (e.g., daily), maintains profiles of normal activity within the enterprise and detects new, suspicious connections to external destinations which are presented to a security analyst for further investigation. Our techniques can work with a number of different datasets that include information on external connections initiated by internal hosts within an organization, for example netflows, DNS logs, full packet captures, web proxies logs, etc.

The first dataset used in this study consists of anonymized DNS logs collected from the Los Alamos National Lab (LANL) internal network. It includes DNS queries initiated by internal hosts, DNS responses from the LANL DNS servers, timestamps of those events, and IP addresses of the sources and destination hosts. All of the IP addresses and domain names are anonymized (consistently --- i.e., the same IP always maps to the same anonymized IP). In addition to real-world DNS logs, the dataset also includes simulated DNS traffic that is representative of those observed during the initial stage of stealthy, targeted attacks (i.e., APTs). A total of 20 distinct attack campaigns are included.
%Logs from the simulated attacks are overlaid unto the legitimate DNS traffic, and LANL issued a challenge problem to the community entitled ``\emph{APT Infection Discovery using DNS Data}'' asking for methods to detect these attacks.

The second dataset \emc\ consists of logs collected by web proxies that intercept HTTP/HTTPs communications at the border of a large enterprise network with over 100,000 hosts. The logs include the connection timestamp, IP addresses of the source and destination, full URL visited, and additional fields specific to HTTP communications (HTTP method, status code, user-agent string, web referer, etc.). As such, the \emc\ dataset is much richer compared to the LANL dataset.

In addition to the web proxy logs, we also obtained access to a list of domain IOCs used by the SOC for detecting malicious activities, and compromised internal hosts communicating to those domains. The IOCs are gathered from external intelligence sources and commercial blacklists, and often serve as the initial point of investigation by SOC analysts. This data was collected at the same time as the web proxies logs.

All of our datasets span an interval of two months (in early 2013 for LANL, and early 2014 for \emc) and are extremely large (1.15TB and 38.14TB, respectively), raising a number of scalability challenges. While the \emc\ dataset is much richer in information, the main advantage of LANL is that it includes the simulated infection campaigns by LANL domain experts and thus provides a labeled dataset by which to validate our methods. Our main results, however, are from evaluations performed on the \emc\ dataset in collaboration with the enterprise SOC.

In the analysis and results presented in the following sections, we focus on ``rare'' destinations in our datasets. Our insight is that popular websites (visited by a large user population) are better administered and less likely to be compromised, but connections to uncommon destinations may be indicative of suspicious behavior. More specifically, we are interested in external destinations that are:
\begin{itemize}
\item \emph{New:} Not visited before by any internal hosts. The rationale is that attackers tend to use new domains under their control for at least some of the attack stages, and also that those malicious domains would not be contacted by benign hosts.

\item \emph{Unpopular:} Visited by a small number of internal hosts. The intuition is that attackers are likely to compromise only a few hosts during initial infection.
\end{itemize}
To identify the ``new'' destinations, we keep track of external destinations contacted by internal hosts over time. This ``history'' of destinations is initialized during a bootstraping period (e.g., one month), and then updated incrementally daily.

These new and unpopular domains are called \emph{rare destinations} and are the starting point of our detection. We found in the enterprise of our study that the number of rare destinations is on the order of 50,000 daily, and the challenge we face is identifying in this set potential malicious domains.

\subsection{Modes of operation}

Our detection method operates in two modes. In the first, called \emph{SOC hints}, we use the incidents that the enterprise SOC investigated as starting points. Given either hosts or domains confirmed malicious by an analyst as seeds, our algorithm identifies other related malicious domains (likely part of the same campaign) and internal compromised hosts that were unknown previously.
%In large enterprises (as the one we study), SOC teams have access to external intelligence sources containing indicators of compromise (IOCs) and commercial blacklists, and often get alerts about internal hosts contacting malicious domains. IOCs are by no means complete, but they can be used as starting points for investigation. The current investigation process by SOC is done manually by human analysts leveraging various log data collected in the enterprise, and additional external resources. We propose a new algorithm inspired by the belief propagation framework from graph theory that aims at automating this process. The algorithm can be seeded with hints provided by SOC (either hosts or domains confirmed malicious by an analyst) and finds other related malicious domains (likely part of the same campaign) and internal compromised hosts.

The second is the \emph{no-hint} mode, in which no known compromised hosts or malicious domains are available. In this mode, we develop a new \cc\ communication detector utilizing connection timing patterns from hosts to domains, domain registration information, and enterprise-specific features extracted from the web proxy logs. Compared to previous work, our \cc\ detector does not require malware samples for training, and can detect \cc\ communication when only one infected host in the enterprise communicates with the external \cc\ domain. Interestingly, the detected \cc\ domains and the hosts contacting them can be used to seed the same belief propagation algorithm and identify additional related suspicious domains and compromised hosts, i.e., serve as input to the \emph{SOC hints} mode.

\ignore{
\vspace{2pt}

\noindent\textbf{SOC hints.} In this case, we start from one host whose compromise is confirmed by the enterprise SOC. Our goal here is to detect the malicious domains controlled by attackers, and other internal hosts that were compromised in the same campaign. The current investigation process by SOC is done manually by human analysts leveraging various log data collected in the enterprise, and additional external resources. The techniques proposed here aim at automating this process.
}

%Since we have access to the incident database in the enterprise of our study including results of human investigation, we have very high confidence that hosts confirmed by a human analyst are in fact compromised.

%\vspace{2pt}

\subsection{Belief propagation framework}

We model the communication between internal hosts and external domains with a bipartite graph $G=(V,E)$, in which there are two types of vertices, hosts and domains. An edge is created between a host and a domain if the host contacts the domain at least once during the observation window (e.g., one day). We would like to label each domain as malicious or benign, and each host as compromised or clean, with high confidence.
%The algorithm is given as input (or seeds) a set of compromised hosts and optionally a set of suspicious domains.

%We're also interested in detecting small communities of domains in this bipartite graphs, domains that are similar with respect to different features, e.g., hosts visiting them, timing patterns, IP space, and likely belong to the same attack campaign.

%The algorithm output can be used to generate these communities of similar domains which likely belong to the same campaign.

%starting with one compromised host $h \in V$ given as hint and no known labels on the domains.

Our main insight is to apply a graph theoretic technique called \emph{belief propagation}~\cite{Pearl82}, commonly used to determine the label of a node given prior knowledge about the node itself and information about its graph neighbors.  The algorithm is based on iterative message-passing between a node and its neighbors until convergence or when a specified stopping condition is met.
For our purposes, the messages (or ``beliefs'') passed between nodes are an indication of suspicious activities observed during early-stage malware infections.
%For our purposes, we adapt the general technique to our problem setting.
In typical implementations, the entire graph is constructed in advance and ``beliefs'' are transmitted from every node to all its neighbors in each iteration. Since the graphs in our case are very large (potentially tens of thousands of domains and hosts daily), we propose an \emph{incremental} method of building the bipartite graph in which hosts and domains are added to the graph only when the confidence of their compromise is high.

\ignore{
Belief propagation is an iterative algorithm which starts with some prior probabilities assigned to each node, called \emph{the node potential function}. In each iteration, a message is transmitted from node $i$ to neighboring node $j$ in the graph denoting intuitively the belief node $i$ has about the label of node $j$. Based on incoming messages from all neighbors, node $j$ generates new beliefs which are transmitted to its neighbors in the next iteration. The algorithm proceeds iteratively until it either converges or some stopping condition is met, at which point the final belief is generated from combining the node's potential function and messages received from neighbors.
}

\ignore{
We build a graph starting from the known compromised hosts (and optionally suspicious domains) given as seeds. In each iteration, we add to the graph and rank the rare domains contacted by known compromised hosts (based on a number of features described in~\secref{sec:cc} and \secref{sec:domsim}) and label the domains with the highest scores as suspicious. The set of compromised hosts is also expanded to include other hosts contacting the newly labeled suspicious domains. The algorithm terminates when the confidence that newly selected domains are malicious is below a certain threshold or when the maximum number of iterations is reached, and returns a list of labeled malicious domains ordered by suspiciousness level.
}

We build a graph starting from the known compromised hosts (and optionally suspicious domains) given as seeds. In each iteration, we compute scores for those rare domains contacted by known compromised hosts, and label the domains with the highest scores as suspicious. The set of compromised hosts is also expanded to include other hosts contacting the newly labeled suspicious domains. The algorithm terminates when the score of the top-ranking domain is below a threshold, or when the maximum number of iterations is reached, and returns a list of labeled malicious domains ordered by suspiciousness level.

The score for a domain is computed based on 1) the degree to which the domain exhibits \cc-like behavior (described in~\secref{sec:ccfeatures}), \emph{or} 2) its similarity to labeled suspicious domains from previous iterations of the belief propagation. In the latter case, the similarity of two domains is based on the overlap in hosts that contacted them, time difference between connections by the same host, and proximity of their IP addresses. These features, combined with properties of the domain itself (domain age, registration validity, popularity of user-agent string and web referer information), are used in a regression model to compute a \emph{similarity score} for the domain relative to the set of domains already labeled suspicious during belief propagation. More details about domain similarity is provided in~\secref{sec:domsim}.

\subsection{Detecting \cc\ communication}
\label{sec:ccfeatures}

As discussed in~\secref{sec:statement} communication to the \cc\ center happens in many campaigns on a regular basis to allow the attacker backdoor access into the compromised system. Such communications are automatically generated by a malicious process on the victim's machine, and exhibit certain regularity in connection timing patterns. By contrast, normal user activities are largely variable.

We leverage these insights to build a novel detector for automated communication that compares the inter-connection histogram of the communication between a host and a domain to that of a periodic (regular) distribution.  The communication between a host and a domain is labeled ``automated'' if the statistical distance between the two histograms is below a certain threshold. Compared to other methods for identifying automated connections (e.g., Fast-Fourier transform in BotFinder~\cite{BotFinder} and autocorrelation in BotSniffer~\cite{BotSniffer}) our method can be tuned for resiliency to outliers and randomization between connections through several parameters. Our detector (like previously proposed timing-based \cc\ detection methods) will miss communications with large variability in timing patterns, but these methods are not commonly used by attackers~\cite{APT1}. Detecting \cc\ communication with completely randomized timing patterns (without access to known malware samples) remains an open problem to the community.

However, thousands of legitimate requests have regular timing patterns as well (due to site refreshes or automatic updates). Restricting our focus to \emph{rare} domains significantly reduces the number of considered domains, but we need to leverage additional features to identify the truly suspicious automated connections. Some of these features have been used in previous work for identifying generic malicious activities (e.g., domain age and registration validity extracted from WHOIS data), and some are tailored to an enterprise environment (e.g., popularity of user-agent strings within that network and web referer information in the HTTP traffic).

Combining these features, we train a linear regression model to output a score for each domain that is detected to receive automated, periodic, communications. More specifically, the score for a domain is a weighted sum of the features, where the weights are determined by the regression model during training. Domains with score higher than a threshold (determined based on tradeoffs between accuracy and coverage) are considered potential \cc\ domains.

\ignore{
To determine the weight of each feature, we use a regression model on a set of training domains with automated connections. The model outputs for each automated domain a score based on the feature weights computed during training. Domains with score higher than a threshold (determined experimentally based on tradeoffs between accuracy and coverage) are considered potential \cc\ domains.
}

%The domain(s) of maximum score (if above a threshold) are labeled malicious, the hosts contacting them are added to the graph and the algorithm proceeds iteratively until the maximum domain score is below a certain threshold.

\subsection{Putting it all together}

\begin{figure*}[t]
%\vspace{-1.2cm}
\centering
\includegraphics[width=5in]{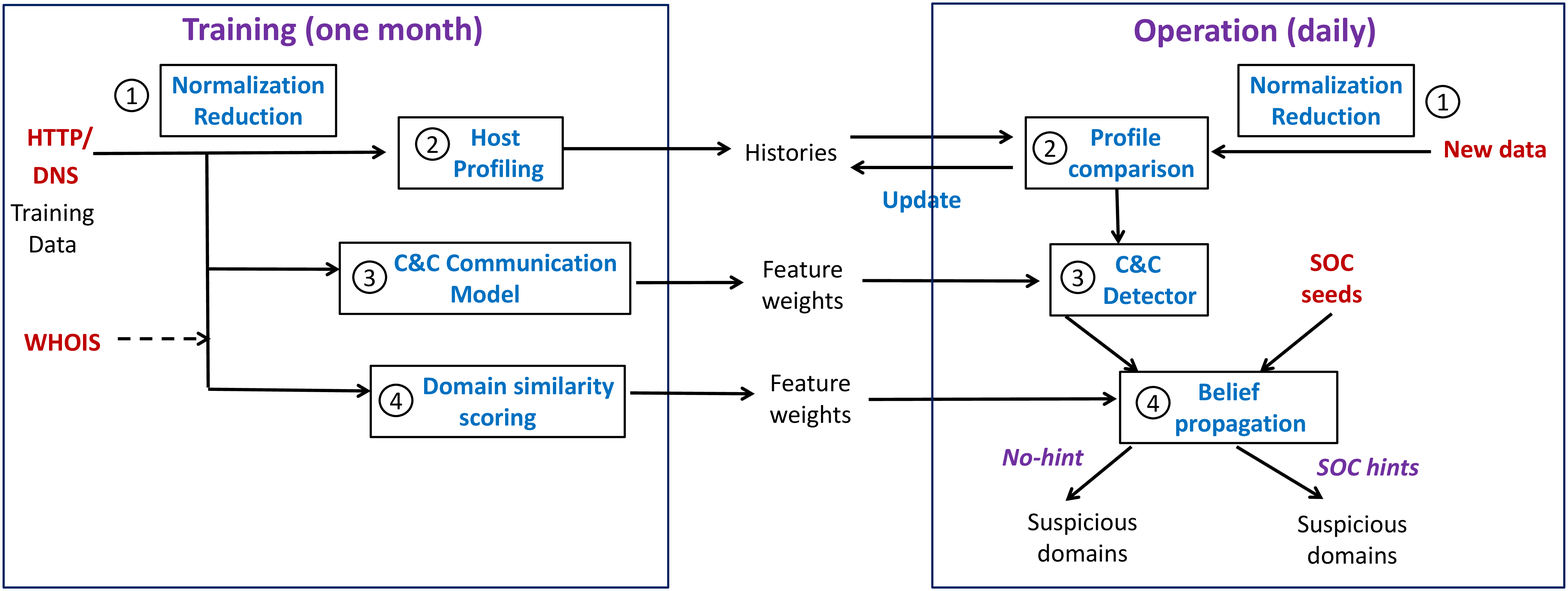}
%\vspace{-0.7cm}
\caption{Overview of training and operation stages in our system for detecting enterprise infection. Training stage is on the left and operation on the right. Input data is shown in red, processing steps in blue and various outputs in black. }
\label{fig:overview}
%\vspace{-0.5cm}
\end{figure*}

Our system for detecting early-stage enterprise infection consists of two main phases: training (during a one-month bootstrapping period) and operation (daily after the training period). An overview diagram is presented in Figure~\ref{fig:overview}.

\vspace{2pt}

\noindent\textbf{Training.} The training period is specific to each organization and its role is to create a benchmark of normal activity of enterprise hosts.  It consists of several steps.

\vspace{1pt}

\noindent (1) \emph{Data normalization and reduction:} The first stage processes the raw log data (either HTTP or DNS logs) used for training and applies normalization and reduction techniques.

\vspace{1pt}

\noindent (2) \emph{Profiling:} Starting from normalized data, the system profiles the activity of internal hosts. It builds histories of external destinations visited by internal hosts as well as user-agent (UA) strings used in HTTP requests (when available).  These histories are maintained and incrementally updated during the operation stage when new data is available.

%Histories of external destinations contacted by internal hosts and UAs used by each host are created.

\vspace{1pt}

\noindent (3) \emph{Customizing the \cc\ detector:} The detector of \cc\ communication is customized to the characteristics of the enterprise. Enterprise-specific features (e.g., rare destinations, popularity of user-agent strings used in a connection, web referer information) are used in combination with other generic features utilized in previous work (automated connections, domain age and validity). A regression model is trained to determine the feature weights for the particular enterprise and the threshold above which a domain is flagged as \cc.

\vspace{1pt}

\noindent (4) \emph{Customizing the domain similarity score:} The domain similarity score used during belief propagation is also customized to the enterprise during the training phase. The weights of the features used for determining domain similarity scores during belief propagation, as well as the score threshold, are also chosen by a regression model.

\vspace{2pt}

\noindent\textbf{Operation.} After the initial training period, the system enters into daily operation mode. Several stages are performed daily:

\vspace{1pt}

\noindent (1) \emph{Data normalization and reduction:} The system processes new data for that day, normalizes it and  performs data reduction.

\vspace{1pt}

\noindent (2) \emph{Profile comparison and update:} New data is compared with historical profiles, and rare destinations, as well as rare UAs (used by a small number of hosts) are identified. Histories of external destinations and UAs are updated with new data, so that changes in normal behavior are captured in the profiles.

\vspace{1pt}

\noindent (3) \emph{\cc \ detector:} The \cc\ detector is run daily, and scores of automated domains are computed with weights determined during training. Automated domains with scores above a threshold are labeled as potential \cc\ domains.

\vspace{1pt}

\noindent (4) \emph{Belief propagation:} The belief propagation algorithm is run in two modes (with or without hints). The output is an ordered list of suspicious domains presented to SOC for further investigation.

\ignore{
\vspace{1pt}

\noindent (9) \emph{Output suspicious domains:} Output ordered list of suspicious domains (in decreasing order of scores) for both modes. Results are presented to SOC for further investigation.
}

\section{System Details}

After providing an overview of our system for detecting early-stage enterprise infection, we give here more technical details of our methods.

\subsection{Data Normalization and Reduction}
\label{sec:datareduction}

\noindent{\bf LANL dataset.} LANL released a dataset comprised of anonymized DNS traffic collected over two months (February and March 2013) from their internal network~\cite{LANL}. The entire LANL dataset consists of 3.81 billion DNS queries and 3.89 billion DNS responses, amounting to 1.15 TB. To allow efficient analysis, we employ a number of data reduction techniques.  We first restrict our analysis only to A records, as they record the queries to domain names and their responses (IP addresses) and information in other records (e.g., TXT) is redacted and thus not useful. This step prunes 30.4\% of DNS records on average per day. We also filter out queries for internal LANL resources (as our focus is on detecting suspicious external communications), and queries initiated by internal servers (since we aim at detecting compromised hosts).

\ignore{IP addresses of internal hosts are anonymized consistently in a prefix-preserving manner. For domains, each level of the name is anonymized separately, mapped to a random character string with the same length. In addition to legitimate traffic, data from the month of March also includes attack traces from 20 independent APT attack campaigns simulated by LANL domain experts. Each simulation is an instance of the initial first-day evolution of an independent campaign.}

\vspace{2pt}

\noindent{\bf \emc\ dataset.} The \emc\ dataset consists of web proxies logs generated at the border of a large enterprise over a period of two months (January and February 2014). Analyzing the \emc\ dataset proved difficult due to its large scale, as well as inconsistent information. There are on average 662GB of log data generated daily, resulting in a total of 38.14TB of data over the two months of our investigation. This dataset is 33 times larger than the LANL dataset, and much richer in information. Compared to LANL data in which all timestamps are in the same time zone and IP addresses are statically assigned, the \emc\ dataset has some inconsistencies due to multiple time zones (as the collection devices are in different geographical locations) and DHCP and VPN assignment for most of the IP address space.

We omit here a detailed description of our normalization procedure, but we converted all timestamps into UTC and DHCP and VPN IP addresses to hostnames (by parsing the DHCP and VPN logs collected by the organization). After normalization, we extract the \textit{timestamp}, \textit{hostname}, \textit{destination domain}, \textit{destination IP}, \textit{user-agent string}, \textit{web referrer} and \textit{HTTP status code} fields for our analysis. We do not consider destinations that are IP addresses.

\begin{figure}[tbhp]
\centering
%\vspace{-0.8cm}
\includegraphics[width=2.5in, angle=270]{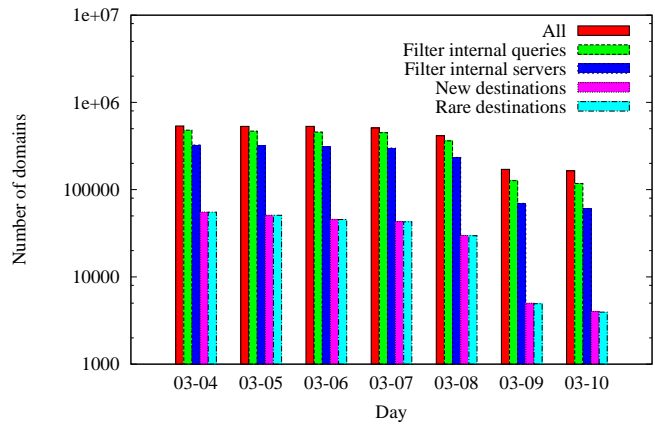}
%\caption{Number of domains after data reduction for a week.}
%\vspace{-0.8cm}
\caption{The number of domains encountered daily in LANL after data reduction for the first week of March.}
%The number of active domains and hosts notably drop on 03-09 and 03-10 since these days are weekend days.
\label{fig:reduction}
%\vspace{-0.4cm}
\end{figure}

\vspace{2pt}

\noindent{\bf Rare destinations.} For both datasets, we use the first month of data for profiling and building a history of external destinations visited by internal hosts. We determine for each day of the second month a list of new destinations not visited before by any internal host.  The rationale is that attackers tend to use new domains under their control not visited previously by internal hosts, and benign hosts (the large majority) are unlikely to visit malicious sites contacted by compromised hosts.

For profiling external destinations, we first ``fold'' the domain names to second-level (e.g., \texttt{news.nbc.com} is folded to \texttt{nbc.com}), assuming that this captures the entity or organization responsible for the domain. Since domain names are anonymized in the LANL dataset, we conservatively fold to third-level domains (as we don't have information on top-level domains). We then maintain a history of (folded) external destinations queried by internal hosts, updated at the end of each day to include all new domains from that day. A domain is considered to be \emph{new} on a particular day if it has not been contacted before, and \emph{unpopular} is it has been queried by less than a certain threshold of distinct hosts in a single day. We set the threshold at 10 hosts based on discussion with security professionals at a large organization. These \emph{rare destinations} are the starting point for our investigation.

Following the steps detailed above, we greatly reduce the size of both datasets. The number of domains after each reduction step in LANL for one week in March is shown in Figure~\ref{fig:reduction}. On average, while the full dataset contains queries from almost 80,000 hosts to more than 400,000 domains per day, in the reduced dataset only 3,369 hosts querying 31,582 domains are included daily on average. In the \emc\ dataset starting from 120K hosts and 600K domains in the original dataset, after data reduction we encounter on average 20K hosts and  59K rare domains daily.

\subsection{Belief Propagation Algorithm}

The belief propagation algorithm can be applied in two modes: with hints of compromised hosts provided by SOC, or without hints. In the first case we use as seed  a list of compromised hosts investigated and confirmed by security analysts in SOC and optionally a list of malicious domains. In the latter case the \cc\ communication detector is run first to identify a set of potential \cc\ domains and hosts contacting them. These are given as seeds to the belief propagation algorithm. The algorithm is run daily in both modes and it detects malicious domains and compromised hosts that are likely part of the same campaign with the provided seeds.

Algorithm 1 presents the pseudocode of the belief propagation algorithm. The algorithm is initialized with the set of compromised hosts $\mathcal{H}$, and set of malicious domains $\mathcal{M}$ (when available). In the \emph{SOC hints} case, $\mathcal{H}$ is the set of hint hosts and $\mathcal{M}$ is the set of malicious domains (if provided). In the \emph{no-hint} case, $\mathcal{M}$ is the set of potential \cc\ domains and $\mathcal{H}$ is the set of hosts contacting them.

The algorithm maintains several variables: $\mathcal{R}$ the set of rare domains contacted by hosts in $\mathcal{H}$ and $\mathcal{N}$ the set of newly labeled malicious domains (in a particular iteration). In each iteration, the algorithm first detects suspicious \cc-like domains among set $\mathcal{R}$ using function \detectcc\ whose exact implementation will be provided next section. Note that in the \emph{no-hint} mode function \detectcc\ will not identify additional \cc\ domains (since they are used for seeding the algorithm and are already included in $\mathcal{M}$). If no suspicious \cc\ domains are found, the algorithm computes a similarity score for all rare domains in $\mathcal{R}$ with function \computescore. The domain of maximum score (if above a certain threshold $\tscore$) is included in set $\mathcal{M}$. Finally the set of compromised hosts is expanded to include other hosts contacting the newly labeled malicious domain(s). The algorithm iterates until the stoping condition is met: either no new domains are labeled as malicious (due to their scores being below the threshold) or the maximum number of iterations has been reached. The output is an expanded lists of compromised hosts $\mathcal{H}$ and malicious domains $\mathcal{M}$.

\begin{algorithm}
\scriptsize
    \begin{algorithmic}
    \State /* $\mathcal{H} \leftarrow \mbox{set of seed hosts}$ */
    \State /* $\mathcal{M} \leftarrow \mbox{set of seed domains}$ */
    \State /* dom\_host is a mapping from a domain to set of hosts contacting it */
    \State /* host\_rdom is a mapping from a host to set of rare domains visited */
    \Function{Belief\_Propagation}{$\mathcal{H}$,$\mathcal{M}$}:
    \State $\mathcal{R} \leftarrow $ set of rare domains contacted by hosts in $\mathcal{H}$
    \While{stop\_condition}
        \State $\mathcal{N} \leftarrow \Phi$ \hspace{0.3cm} /* set of newly labeled malicious domains */
        \For{dom in $\mathcal{R} \setminus \mathcal{M}$}
            \If{\detectcc(dom)}
                \State $\mathcal{N} \leftarrow \mathcal{N} \cup \{dom\}$
                \State $\mathcal{R} \leftarrow \mathcal{R} \setminus \{dom\}$
            \EndIf
        \EndFor
        \If{$\mathcal{N} = \Phi$}
            \For{dom in $\mathcal{R} \setminus \mathcal{M}$}
                \State score[dom] $\leftarrow$ \computescore(dom)
            \EndFor
            \State max\_score $\leftarrow$ max[score[dom]]
            \State max\_dom $\leftarrow$ dom of maximum score
            \If{max\_score $ \ge \tscore$}
                \State $\mathcal{N} \leftarrow \mathcal{N} \cup \{dom\}$
            \EndIf
        \EndIf
        \If{$\mathcal{N} \neq \Phi$}
            \State $\mathcal{M} \leftarrow \mathcal{M} \cup \mathcal{N}$
            \State $\mathcal{H} \leftarrow \mathcal{H} \cup (\cup_{d \in \mathcal{N}}$ dom\_host[d])
            \State $\mathcal{R} \leftarrow \mathcal{R} \cup  (\cup_{h \in \mathcal{H}}$ host\_rdom[h])
        \EndIf
    \EndWhile
    \EndFunction
    \caption{[Belief Propagation]}
    \label{alg:BP}
    \end{algorithmic}
\end{algorithm}

\subsection{Detection of \cc\ communication}
\label{sec:cc}

\vspace{2pt}
\noindent{\bf Dynamic histograms.} We aim at detecting automated connections with fairly regular timing patters, but be resilient to outliers (for instance large gaps in communication) and small amounts of randomization introduced by attackers between connections.

Initially, we tested a detector for automated connections based on standard deviation (labeling the connections between a host and a domain as automated if the standard deviation of the inter-connection intervals is small), but found out that a single outlier could result in high standard deviation. Our main idea is to generate a histogram of inter-connection intervals between a host and a domain,
and compare it to that of a periodic distribution using a known statistical distance. However, the distance metric is highly sensitive to the histogram bin size and alignment. Setting the size to a large value leads to over smoothing, while a small value increases the sampling error. When bins are statically defined, relatively close values might fall under different bins, affecting the distance metric.

We propose  a \emph{dynamic histogram} binning method instead. Here we first cluster the inter-connection intervals (denoted $t_1,\dots,t_m$) of successive connections from a host to a domain on a particular day, and then define the bins dynamically from the generated clusters.
Let the first interval $t_1$ be the first cluster ``hub.'' An interval $t_i$ is considered as part of a cluster if it is within $W$ of the cluster hub. Otherwise, a new cluster with hub $t_i$ is created. $W$ is a fixed value, and acts as our ``bin width.''
This dynamic binning method allows us to accommodate timing randomizations typically introduced by attackers between connections.

Each resulting cluster is considered a bin, its frequency is computed and the resulting histogram is compared to that of the periodic distribution with period equal to the highest-frequency cluster hub. We label the communications between a host and a domain to be automated if their inter-connection histogram is ``close'' to periodic (i.e., within a threshold $J_T$), as determined by the Jeffrey divergence.
For two histograms $H = [(b_i,h_i)]$ and $K=[(b_i,k_i)]$, if $m_i = (h_i + k_i)/2$ the Jeffrey divergence is defined as~\cite{Rubner00}:

%
%\vspace{-0.3cm}
$$d_J(H,K) = \sum_i (h_i \log \frac{h_i}{m_i} + k_i \log \frac{k_i}{m_i}) $$
%\vspace{-0.3cm}

We choose the Jeffrey divergence metric motivated by the fact that it is ``numerically stable and robust to noise and size of histogram bins''~\cite{Rubner00}. We experimented with other statistical metrics (e.g., L1 distance), but the results were very similar and omit them herein. \ignore{We believe that the bin definition (static versus dynamic and bin width) has larger impact on comparing two histograms than the exact bin-by-bin measure. }The bin width $W$ and distance threshold $J_T$ parameters control the resiliency of the method to outliers and randomization introduced by attackers. We discuss their selection according to the LANL dataset in~\secref{sec:params_lanl}.

\vspace{2pt}
\noindent{\bf Additional features.} For each rare automated domain visited by an internal host we extract six additional features which will be used for the \cc\ detector:

\vspace{2pt}

\noindent \emph{Domain connectivity features:} We consider the number of hosts contacting the domain (\nohosts) called \emph{domain connectivity}  and the number of hosts with automated connections to the domain (\noautohosts). The intuition here is that most rare domains (legitimate ones) are contacted by only one host, but those rare domains contacted by multiple hosts are more suspicious as they might indicate multiple compromised hosts under attacker's control.

\vspace{2pt}

\noindent \emph{Web connection features:} Based on discussions with security professional at a large organization, web connections with no web referrer information are more suspicious, as they are not part of a user browsing session and might be generated by a process running on the user machine. Typical user browsing sessions include web referer information, but in some cases the web-referer is wiped out (e.g., java script or iframes embedded in web pages). To capture this, we include a feature \noreferrer\ denoting the fraction of hosts (among all hosts contacting that domain) that use no web referer.

Additionally, most users have a fairly small number of user-agent strings in their HTTP connections (on average between 7 and 9 per user). Software configurations in an enterprise are more homogenous than in other networks (e.g., university campus), and as such we'd expect that most user-agent strings are employed by a large population of users. With this intuition, the \emph{rare user-agent strings}, those used by a small number of hosts, might indicate unpopular software installed on the user machine which can potentially be associated with suspicious activities. We consider a feature \rareua\ denoting the fraction of hosts that use no UA or a rare UA  when contacting the domain.

To determine the popularity of UA strings, we maintain a history of UAs encountered across time and the hosts using those UAs. The UA history is built during the training phase for a period of one month and then updated daily based on new ingested data. An UA is considered rare (after the training period of one month) if it is used by less than a threshold of hosts (set at 10 based on SOC recommendation).

\vspace{2pt}

\noindent \emph{Registration data features:} Attacker-controlled sites tend to use more recently registered domains than legitimate ones~\cite{MaKDD09}. In addition, attackers register their domains for shorter periods of time to minimize their costs in case the campaign is detected and taken down. We query WHOIS information and extract two features: \domainage\ (number of days since the domain was registered), and \domainval\ (number of days until the registration expires).

\vspace{0.2cm}

\noindent {\bf Scoring automated domains.} During the training stage, for each rare domain identified as having automated connections, we extract the six features explained above. We also query VirusTotal for each rare automated domain and label it ``reported'' if at least one anti-virus engine reports it  and ``legitimate'' otherwise. Using the set of automated domains visited in the enterprise for two weeks, we train a linear regression model, implemented using the function {\bf lm} in the R package. The regression model outputs a weight for each feature, as well as the significance of that feature. The final score for each automated domain is a linear combination of feature values weighted by regression coefficients. Finally, based on the model, we select a score threshold $T_c$ above which domains are labeled as potential \cc\ domains. We provide results on the regression model and threshold selection in Section~\ref{sec:params_emc}.

During the operation stage, domain scores are computed using the weights of the regression model (built during training). Function \detectcc\ in Algorithm 1 returns true for a domain if automated connections are detected to the domain (from at least one host) and the domain score is above the threshold $T_c$ computed during training.

\subsection{Domain similarity}
\label{sec:domsim}

We consider a number of features when computing similarity of a domain $D$ with a set of domains $S$ labeled malicious in previous iterations of belief propagation.

\vspace{1pt}

\noindent \emph{Domain connectivity.} First is the domain connectivity as defined above.

\vspace{2pt}

\noindent \emph{Timing correlations.} Second, we consider features related to the time when the domain $D$ was visited by internal hosts. During initial infection stage of a campaign, we suspect that a host visits several domains under the attacker's control within a relatively short time period (as explained in \secref{sec:statement}). We thus consider the minimum timing difference between a host visit to domain $D$ and other malicious domains in set $S$. The shorter this interval, the more suspicious $D$ is.

%For example, clicking on embedded links in spear-phishing emails directs the host to the front-end attacker site, which further redirects to another site hosting the malicious payload. After the payload successfully executes, the host reaches out to the \cc\ server in another domain shortly.

\vspace{2pt}

\noindent \emph{IP space proximity.} Third, we consider proximity in IP space between $D$ and malicious domains in set $S$, i.e., if they are in the same IP/16 or IP/24. The intuition here is that attackers host a large number of malicious domains under a small number of IP subnets (i.e., IP/24 or IP/16 subnets)~\cite{HaoIMC11, APT1}.

\vspace{2pt}

We provide measurement of the timing and IP proximity features on the LANL dataset in~\secref{sec:params_lanl}. In addition, when computing domain scores in belief propagation some of the features introduced for scoring \cc\ domains (\noreferrer, \rareua, \domainage, \domainval) are also used, as they might indicate suspicious activities.

During the training stage we build a linear regression model with eight features in total, using a set of (non-automated) rare domains visited over two weeks. Results and threshold selection are provided in Section~\ref{sec:params_emc}. During the operation stage, function \computescore\ in Algorithm 1 computes a domain similarity score by using the feature weights found by this model relative to the set of already labeled malicious domains in previous iterations.

\section{Evaluation on the LANL Dataset}
\label{sec:lanl}

We start by describing the four cases in the LANL challenge problem. Then we discuss how we adapted our techniques developed for enterprise infection to detecting the LANL attacks (when less information is available about domain names and connection information). Still, using less features, we are able to demonstrate that our belief propagation framework achieves excellent results on the LANL challenge problem.

\subsection{The LANL Challenge Problem}

The LANL dataset includes attack traces from 20 independent infection campaigns simulated by LANL domain experts. Each simulation is an instance of the initial first-day infection stage  of an independent campaign. LANL issued the \emph{APT Infection Discovery Challenge} to the community requesting novel methods for the detection of malicious domains and compromised hosts involved in these attack instances~\cite{LANL}.  More specifically, each of the simulated attacks belongs to one of four cases in increasing order of difficulty, described in Table~\ref{tab:cases}. Cases 1-3 include ``hints'' about the identity of one or multiple compromised hosts, while no hint is available for case 4. Answers (i.e., the malicious domains) are also given for each attack for validation.

\begin{table}
\scriptsize
\begin{center}
%\centering
\begin{tabular}{|c|l|l|l|}
\hline
Case & Description & Dates & Hint Hosts   \\
\hline  \hline
1 & From one hint host detect the & 3/2, 3/3, 3/4, & One per day  \\
  & contacted malicious domains. & 3/9, 3/10 &   \\
\hline
2 & From a set of hint hosts detect & 3/5, 3/6, 3/7, 3/8, & Three or four \\
  & the contacted malicious domains. & 3/11, 3/12, 3/13 & per day  \\
\hline
3 & From one hint host detect the   & 3/14, 3/15, 3/17,  & One per day \\
  & contacted malicious domains and  & 3/18, 3/19, 3/20,  & \\
  & other compromised hosts. & 3/21 & \\
\hline
4 & Detect malicious domains and & 3/22 & No hints \\
 & compromised hosts without hint. & & \\
\hline
\end{tabular}
\end{center}
%\vspace{-0.3cm}
\caption{The four cases in LANL challenge problem.}
%\vspace{-0.8cm}
\label{tab:cases}
\end{table}

\subsection{Parameter selection}
\label{sec:params_lanl}

When selecting various parameters for our algorithms, we separate the 20 simulated attacks into two equal-size sets, and use one for training (attacks from 3/2, 3/3, 3/4, 3/5, 3/7, 3/12, 3/14, 3/15, 3/17, and 3/18), and the other for testing. We try to include attacks from each case in both training and testing sets, with the only exception of case 4, simulated only on one day (3/22). We deliberately add this most challenging attack (in which no hint is provided) to the testing set. We use the training set for selecting parameters needed for different components of the algorithm (dynamic histogram method, \cc\ communication detection, features used for domain scoring, etc.). We show that parameters chosen according to the training set perform well on new data (testing set).

\begin{table}[th]
\centering
\scriptsize
\begin{tabular}{|c|c|c|c|c|}
\hline
Bin width & Jeffrey distance &  Malicious & Malicious & All automated \\
$W$ & threshold  & pairs in & pairs in & pairs in \\
& $J_T$ & training & testing & testing days \\
\hline
5 seconds & 0.0 & 12 & 15 & 7495 \\
& 0.034  &  14 &  15 & 8070 \\
& 0.06 & 15 & 17 & 8579 \\
& 0.35 & 15 & 18 & 34719 \\
\hline
10 seconds & 0.0 & 12 & 16  & 15611 \\
 & 0.034 & 14  & 16  & 16224  \\
 & 0.06 & 15  & 18  & 16803  \\
\hline
20 seconds & 0.0 & 12  & 15 & 23352 \\
 & 0.034 & 14  & 16  & 23964 \\
 & 0.06 & 15 & 18 & 24597  \\
\hline
\end{tabular}
%\vspace{-0.1cm}
\caption{Number of automated malicious (host, domain) pairs in training and testing sets, as well as the number of automated pairs for all days in testing set.}
\label{tab:beacon}
%\vspace{-0.8cm}
\end{table}

\vspace{2pt}

\noindent \textbf{Thresholds for dynamic histograms.} The dynamic histogram method compares the histogram of inter-connection intervals from successive connections by a host to a domain on a particular day to that of a periodic distribution with the goal of identifying automated communications. As described in~\secref{sec:cc} the method can be configured with two parameters: bin width ($W$) denoting the maximum distance between the bin hub and other intervals in the same bin, and the threshold ($J_T$) denoting the maximum Jeffrey distance between the two histograms. A connection with histogram at distance less  than $J_T$ from the periodic histogram is considered automated. Intuitively, the larger $W$ and $J_T$, the more resilience the method provides against randomization and outliers, but more legitimate connections are labeled automated as well.

%which control how resilient the detector is against outliers (gaps in communication) and randomization introduced by attackers in inter-connection intervals.

To determine the most suitable parameterization, we experiment with 3 different bin widths (5, 10 and 20 seconds) and choose the distance threshold $J_T$ according to the training set of malicious automated connections. We manually labeled as automated 15 (host, domain) pairs in training set and 18 pairs in testing set corresponding to 18 distinct domains.\ignore{ (containing 15 (host, domain) pairs).} Table~\ref{tab:beacon} shows the number of malicious (host, domain) pairs labeled automated (in both training and testing sets), as well as all pairs labeled automated in the testing days for several choices of $W$ and $J_T$. Intuitively, fixing one of the parameters $W$ or $J_T$ and increasing the other results in more legitimate domains labeled automated.

%Intuitively, for a fixed $W$ increasing $J_T$ results in more legitimate domains labeled automated. When fixing $J_T$, increasing $W$ also results in more legitimate domains labeled automated.

For our purposes, we aim at capturing all malicious pairs in the training and testing sets, while labeling fewest legitimate connections automated. For the 10 and 20-second bin width, the threshold capturing all 33 malicious pairs is 0.06, while for the 5-second bin width we need to increase the threshold at 0.35. This larger threshold has the effect of increasing the number of legitimate pairs labeled automated. Based on these results, we choose a bin size of 10 seconds and a threshold distance of 0.06 to achieve our desired properties.

\vspace{2pt}

\noindent \textbf{Features.} Since domain names in the LANL dataset are anonymized and the data contains only DNS requests, we have access to a smaller number of features than in the enterprise case. Features related to domain registration (domain age and registration validity), and features extracted from HTTP connections are not available in LANL.

For detecting \cc\ communication, there are thousands of automated domains daily (up to 5239). Restricting to rare domains is beneficial in reducing the number of automated domains by a factor of more than 100, but we still observe hundreds of rare automated domains. To identify \cc\ communications among this set, we combine multiple features that are available in this dataset, in particular domain connectivity and similarity in timing patterns across hosts. The \cc\ detector for LANL is very simple: we consider an automated domain as potential \cc\ if there are at least two distinct hosts communicating with the domain at similar time periods (within 10 seconds). This heuristic works well because in the LANL simulations there are always multiple infected hosts in every campaign. However, in our most general \cc\ detector for the enterprise case, we consider domain connectivity as a feature which can be combined with other features extracted from registration data and HTTP connection. Our general method can detect \cc\ domains contacted by a single host.

\begin{figure}[t]
%\vspace{-0.3cm}
\begin{center}
  \begin{tabular}{lr}
  \includegraphics[width=2.5in,angle=270]{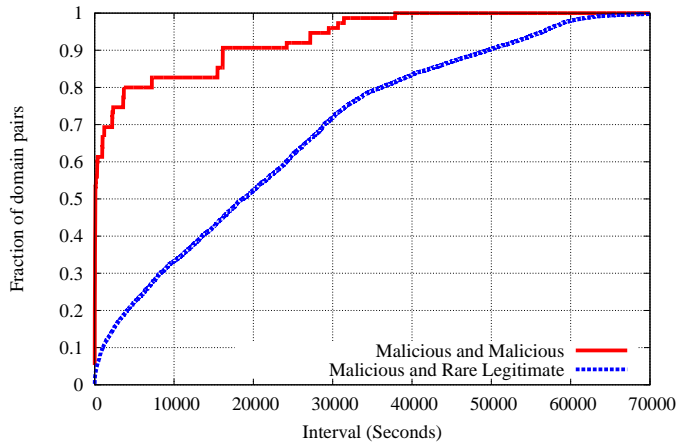}
  \end{tabular}
\end{center}
%\vspace{-0.8cm}
\caption{The CDFs between first connection to two malicious domains and a malicious and legitimate domain by a host.}
%\vspace{-.8cm}
\label{fig:timingcdf}
\end{figure}

Using the LANL simulated attacks, we'd like to measure the relevance of the timing and IP space similarity features among malicious domains. For compromised hosts in the training set, we extract the timestamp of their first connection to every rare domain visited. We plot in Figure~\ref{fig:timingcdf} CDFs of the distributions of the time difference between visits to malicious domains and a legitimate and malicious domain by the same host. The graph confirms that connection intervals between two malicious domains are much shorter than between a malicious and a legitimate domain. For example, 56\% of visits to two malicious domains happen at intervals smaller than 160 seconds, while only 3.8\% of malicious-to-legitimate connection intervals are below this threshold (similar results are observed on testing dataset).

%This demonstrates that timing difference between connections to two domains by the same host is clearly a distinguishing feature for malicious domains.

Next we measure similarity in IP space for malicious and legitimate domains in the training set. We found that 7 malicious domain pairs are in the same IP/24 subnet, while 18 share an IP/16 subnet. We observed few cases of legitimate domains residing in the same subnet with malicious ones. With the exception of 3/7, when more than 2000 pairs of malicious and legitimate domains share the same IP24 or IP16 subnet (due to a single malicious domain belonging to a popular service), the rest of days we observe 20 pairs in the same IP24 subnet and 155 pairs in the same IP16 subnet. We thus use both IP/16 and IP/24 space features in detection but with different weights.

\vspace{2pt}

\noindent \textbf{Domain similarity scores.} In a particular iteration of belief propagation a domain $D$ is compared to the set of domains $S$ already labeled malicious in previous iterations. The domain score is computed as a function of three components: domain connectivity, timing correlation with a known malicious domain in $S$ (value 1 if the domain is contacted close in time to a malicious domain and 0 otherwise), proximity in the IP space with malicious domains in $\mathcal{M}$ (value 2 if same /24 subnet with a malicious domain, 1 if same /16 subnet with a malicious domain and 0 otherwise). Each of these components increases the domain's score. While in our general framework, we proposed a linear regression model for computing domain scores (see~\secref{sec:cc} and \secref{sec:domsim}), we cannot apply that technique here due to limited samples of training data. Instead, we choose a simple additive function that computes the score as the sum of the three components above and then normalizes it. This performs well in both cases 3 and 4 (based on the training set we set the domain score threshold $\tscore$ at 0.25).

\subsection{Results}
\label{sec:lanlres}

We omit here description of the first two cases due to space limitations.

\vspace{2pt}

\noindent\textbf{Starting from a hint host (case 3).} We ran the belief propagation algorithm for a maximum of five iterations starting from the provided hint host, but we stop the algorithm if the maximum domain score is below the threshold $\tscore$.

\begin{figure}[t]
%\vspace{-1.2cm}
\centering
\includegraphics[width = 3in]{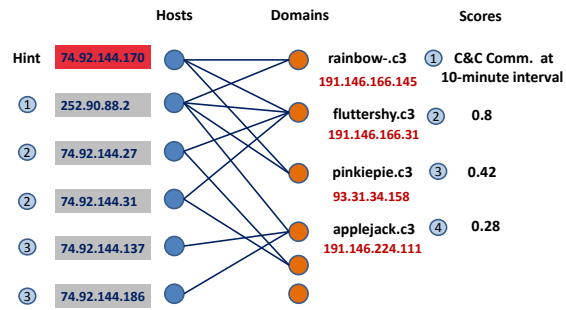}
%\vspace{-0.7cm}
\caption{Application of belief propagation to the 3/19 campaign.}
\label{fig:bpexample}
%\vspace{-0.8cm}
\end{figure}

An example of applying the algorithm to detecting the campaign on 3/19 in the LANL dataset is given in Figure~\ref{fig:bpexample}. Starting from hint host 74.92.144.170, in the first iteration \cc\  communication to domain {\small \tt rainbow-.c3} is detected at 10-minute average intervals. The domain is labeled malicious and the second host (252.90.88.2) compromised. In the following three iterations, three rare domains of maximum score are labeled malicious based on similarity with previously labeled malicious domains (their scores are given on the right). The algorithm stops at the fifth iteration when the maximum score is below a threshold, at which point all labeled domains are confirmed malicious.

The algorithm detects all 12 labeled malicious domains in training days (with no false positives) and all 12 labeled malicious domains in testing days at the cost of one false positive for 3/21.

\vspace{2pt}

\noindent\textbf{No hints (case 4).} In the most challenging case  no hint on compromised hosts are available to seed the belief propagation algorithm. We thus identify first \cc\ communication and seed the belief propagation algorithm with the \cc\ domains. Interestingly, the same algorithm from case 3 delivered very good results on case 4, where we did not have an opportunity for training (case 4 was simulated only on a single day). The five domains identified by belief propagation were confirmed malicious (by the simulation answer), and the algorithm did not have any false positives.

%Among all 25,166 rare domains on 3/22, there are 142 domains with automated activities. Among these, only two domains have at least two hosts beacon at similar intervals: {\tt \small otyugh.muck.don} and {\tt \small nalfeshnee.muck.don}.

%We label these two domains malicious and seed them in the belief propagation algorithm. The four hosts contacting these domains are also given as seed. After three iterations, three other rare domains of maximum score are labeled as malicious, and the algorithm stops in the fourth iteration as the maximum domain score is 0.22.

\vspace{2pt}

\noindent\textbf{Summary.} The summary of our results on the four cases of the LANL challenge are given in Table~\ref{tab:lanlres}. We define several metrics of interest: \emph{true detection rate} (TDR) is the fraction of true positives  and \emph{false detection rate} (FDR) is the fraction of false positives among all detected domains; and \emph{false negative rate} (FNR) is the fraction of malicious domains labeled as legitimate by our detector. Overall, we achieve TDR of 98.33\% (97.06\% on the testing set), with FDR of 1.67\% (2.94\% on testing set) and FNR of 6.35\% (2.94\% on the testing set).

\begin{table}[ht]
\centering
\scriptsize
\begin{tabular}{|c|c|c|c|c|c|c|}
\hline
Case &  \multicolumn{2}{|c|}{True Positives} & \multicolumn{2}{|c|}{False Positives} & \multicolumn{2}{|c|}{False Negatives} \\
\cline{2-7}
& Training & Testing  & Training & Testing & Training & Testing \\
\hline
Case 1 & 6 & 4 & 0 & 0 & 2 & 0 \\
Case 2 & 8 & 12 & 0 & 0 & 1 & 1 \\
Case 3 & 12 & 12 & 0 & 1 & 0 & 0 \\
Case 4 &  - & 5 & - & 0 & - & 0  \\
\hline
Total & 26 & 33  & 0 & 1 & 3 & 1 \\
\hline
\end{tabular}
\caption{Results on LANL challenge.}
\label{tab:lanlres}
\end{table}

\section{Evaluation on Enterprise Data}
\label{sec:eval_emc}

We implemented a fully operational system running in production starting from January 1 2014 to process the web proxies logs from the \emc\ dataset. We use the data collected in January for training various components of the system (e.g., the \cc\ detector, the domain scoring module, etc.) and profiling external destinations and user-agent strings used by enterprise hosts in HTTP communication. Starting from February 1 the system enters into the daily operation mode, in which it processes new web proxies logs, applies normalization and reduction techniques, compares the data with the profiles (which are also updated) and applies our detection techniques. First \cc\ communication is identified, and domains labeled as \cc\ are given as seeds to belief propagation. Second, domains confirmed by SOC (if available) and hosts contacting them are provided as seeds in belief propagation. The detection results are thoroughly validated through a combinations of tools and manual analysis.  The system is configurable with different parameters (e.g., scoring thresholds, number of iterations in belief propagation, etc.) according to the SOC's processing capacity. We present the parameter selection, our validation methodology and the results in both modes of operation.

\ignore{
We implemented a fully operational system running in production starting from January 1 2014 to process \emc dataset. We use the data collected in January for profiling and building the history of visited external destinations. Starting from February 1 the system runs the two detection algorithms (with and without hints) daily. For each run, we set different thresholds for detecting \cc\ communication and belief propagation to examine the effectiveness under different settings. We run belief propagation for a maximum number of 20 iterations, and output at most 40 domains daily. The detection results are then thoroughly validated through a combinations of tools and manual analysis. The thresholds used here are configurable according to the SOC's processing capacity.
%The alerts generated (which include both suspicious domains and internal hosts) are presented to a security analyst for further investigation.}
}

\subsection{Parameter selection}

\label{sec:params_emc}

\noindent {\bf Scoring automated domains.} The parameters of the dynamic histogram method selected in~\secref{sec:params_lanl} result in 841 distinct automated domains for the month of February in the \emc dataset. We query VirusTotal for each rare automated domain and label it ``reported'' if at least one anti-virus engine reports it  and ``legitimate'' otherwise (144 domains were reported by VirusTotal). We thus generate a large set of labeled automated domains and use it to learn the relevance of each feature in identifying suspicious \cc\ domains.

In total, we observe 2888 automated (host, domain) pairs in the month of February for which we can extract all features described in~\secref{sec:cc}. We divide the labeled set of automated domains into a training set covering the first two weeks of February, and testing set for the last two weeks. We construct a linear regression model on the training set, which outputs a coefficient for each feature, as well as the significance of that feature. Among all features considered, the only one with low significance was \noautohosts, which we believe is highly correlated with \nohosts\ and thus omit it. The most relevant features found by the model are \domainage\ and \rareua. \domainage\ is the only one negatively correlated with reported domains (as they are in general more recently registered than legitimate ones), but all other features are positively correlated.

\begin{figure}[tp]
%\vspace{-0.8cm}
\begin{center}
  \begin{tabular}{lr}
  \includegraphics[width=2.5in,angle=270]{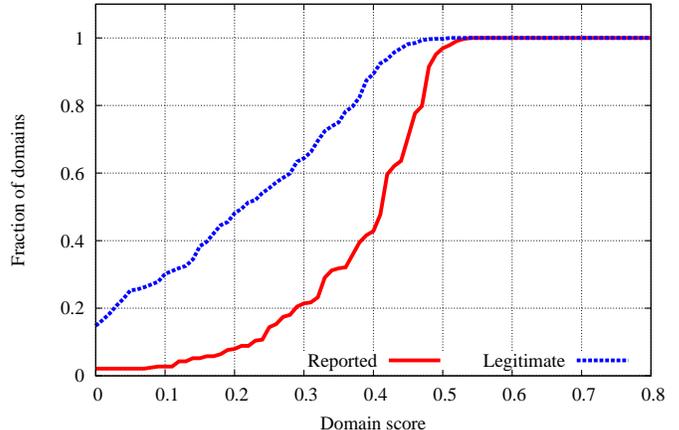}
  \end{tabular}
\end{center}
%\vspace{-0.8cm}
\caption{CDFs of automated reported/legitimate domain scores.}
%\vspace{-0.4cm}
\label{fig:sbeacon}
\end{figure}

The graph in Figure~\ref{fig:sbeacon} shows the difference between the scores of automated domains reported by VirusTotal and legitimate ones on the training set. We observe that reported domains have higher scores than legitimate ones. For instance, selecting a threshold of 0.4 for labeling an automated domain suspicious results in 57.18\% true detection rate and 10.59\% false positive rate on the training set, and 54.95\% true detection rate and 11.52\% false positives on the testing set. We emphasize that our final goal is not identifying {\bf all} automated domains reported by VirusTotal, but rather a significant fraction that allow us to bootstrap the belief propagation algorithm and find new suspicious domains not yet detected by current anti-virus technologies.

\vspace{2pt}

\noindent {\bf Domain similarity.} To obtain a list of (non-automated) rare domains and their features, we start from a set of compromised hosts (those contacting the \cc\ domains confirmed by VirusTotal). We include each rare domain contacted by at least one host in this set, extract its features, query VirusTotal to get an indication of its status, and divide the data into training and testing set, covering the first and last two weeks of February, respectively.

We apply again linear regression on the training set to determine feature weights and significance. Among the eight considered features described in~\secref{sec:domsim}, the only one with low significance was \ipsixteen, as we believe it's highly correlated with \iptwofour. The most relevant features identified by the model are \rareua, \dominterval, \iptwofour\ and \domainage. We still observe a difference in the score CDFs for reported and legitimate domains (we omit the plot due to space limitations).

%For instance, for a threshold of 0.33 we get a true detection of 34.07\% and false positive rate of 10.52\% on the training set and 40.8\% true detection and 11.37\% false positive rate on the testing set.

%We configure several parameters in order to bound the number of detected domains daily. We run belief propagation for a maximum number of 20 iterations, and output at most 40 domains daily. We also set thresholds on domain scores, and only output domains that have higher score than the selected thresholds. The thresholds are configurable according to the capacity of the team performing manual investigation.}

%We present results here only for the month of February, but the system is run daily on March and April as well.

\begin{figure*}[tbhp]
\centering
%\vspace{-0.8cm}
\begin{subfigure}[b]{0.3\textwidth}
\centering
\includegraphics[width=2.2in]{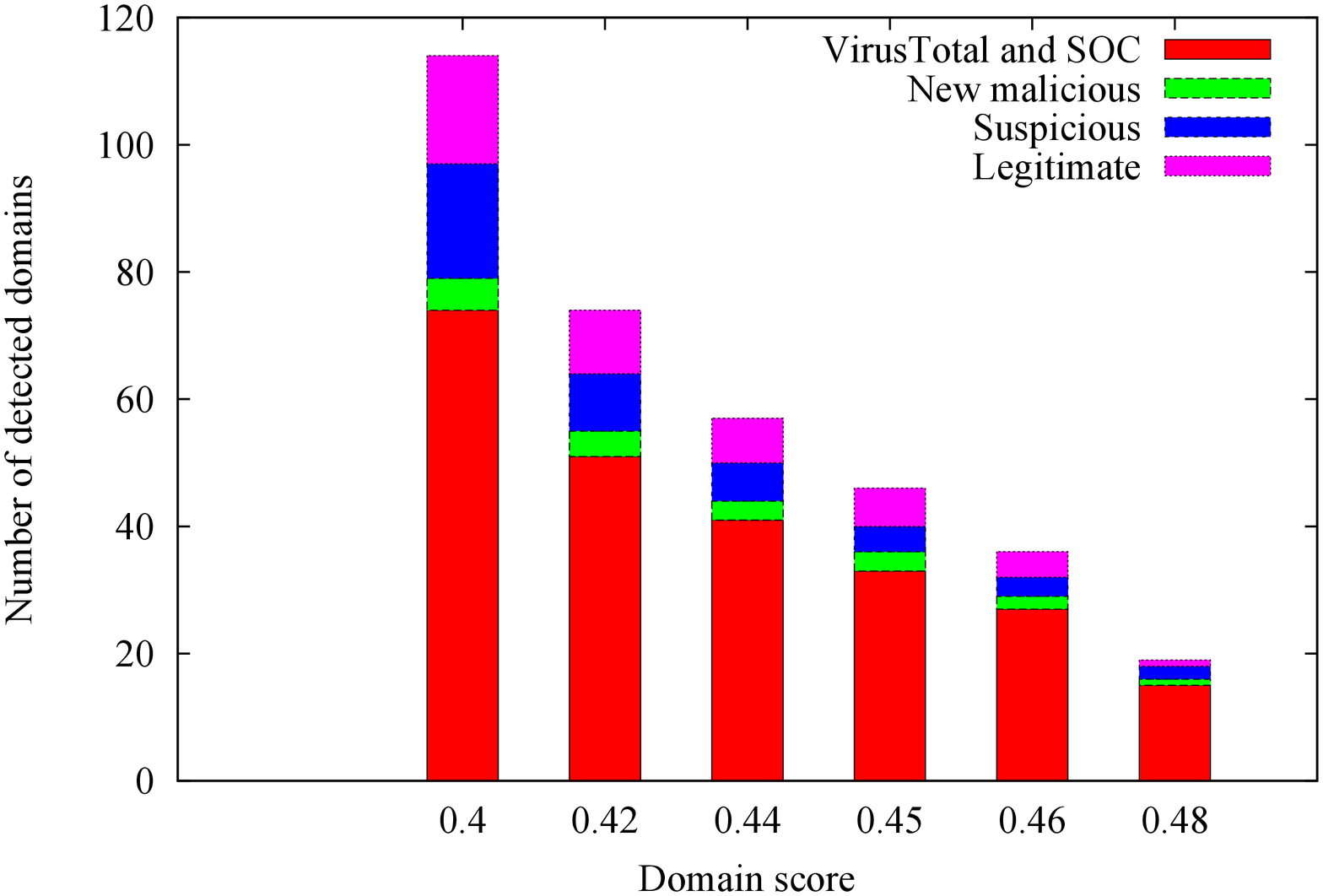}
\end{subfigure}
\begin{subfigure}[b]{0.3\textwidth}
\centering
\includegraphics[width=2in]{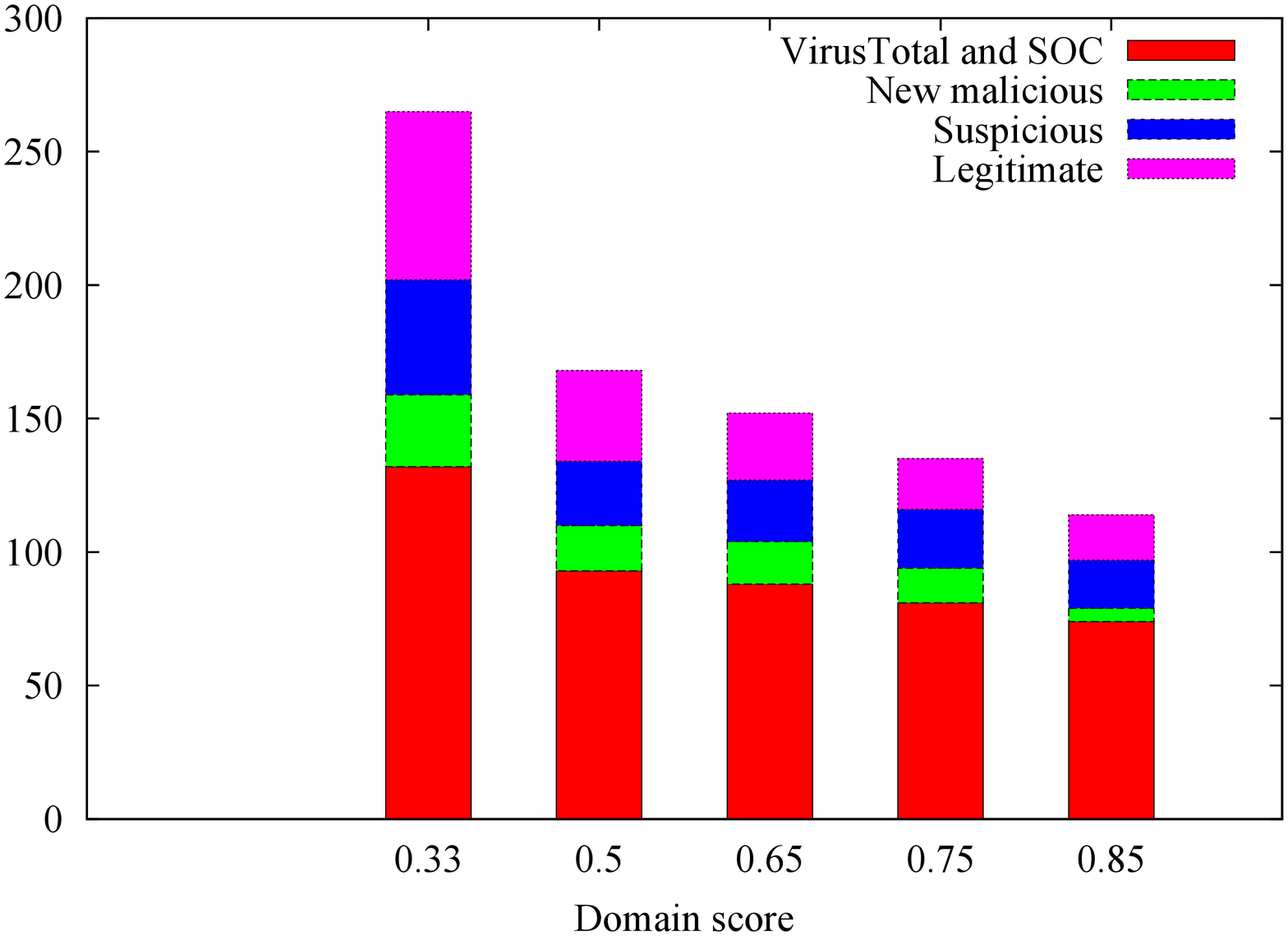}
\end{subfigure}
\begin{subfigure}[b]{0.3\textwidth}
\centering
\includegraphics[width=2in]{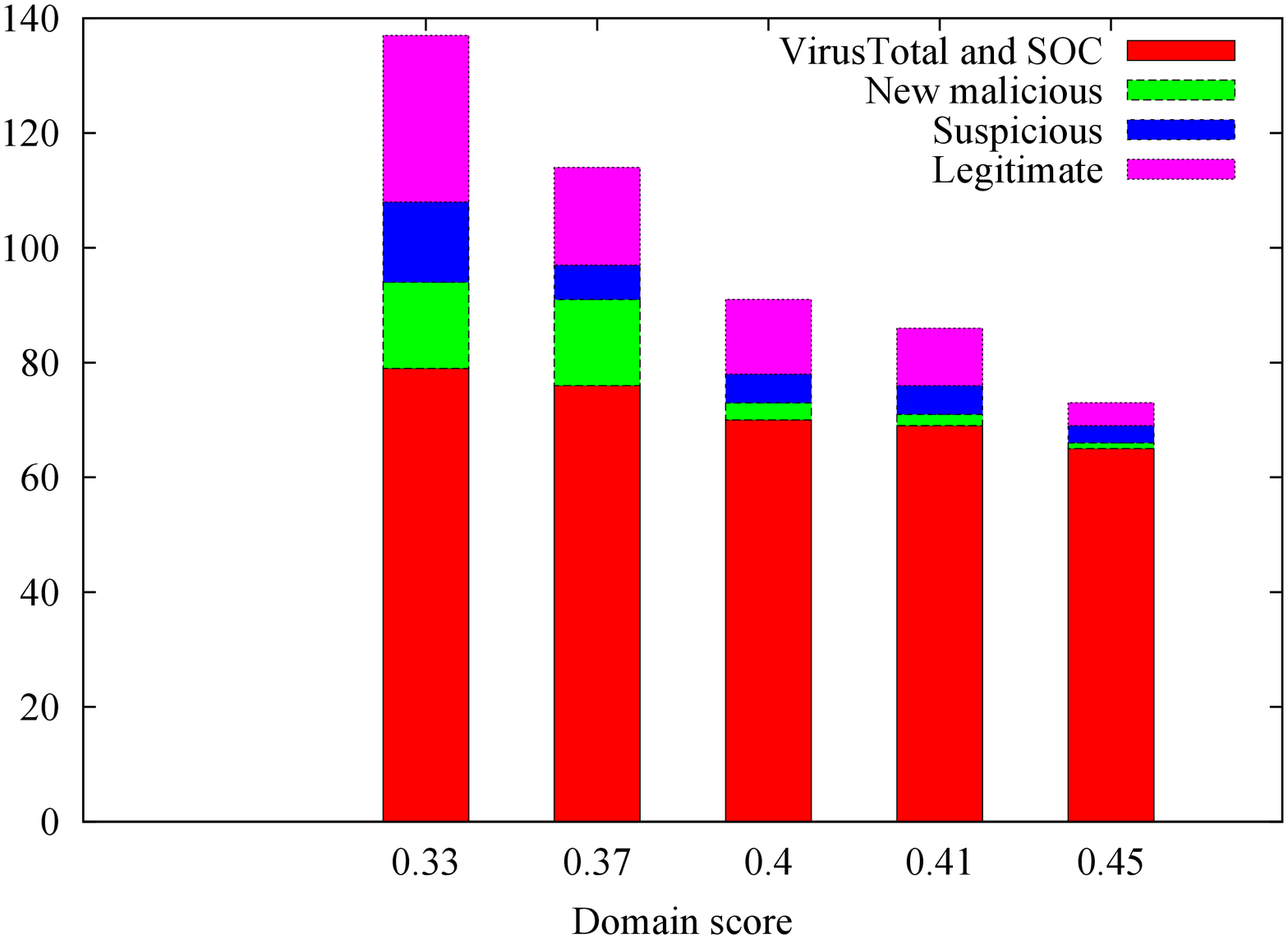}
\end{subfigure}
%\vspace{-0.5cm}
\caption{Statistics on detected domains. (a) \cc\ communication. (b) No hints. (c) SOC hints.}
\label{fig:emceval}
%\vspace{-0.5cm}
\end{figure*}

\subsection{Validation methodology}

We ran the detector in both modes, \emph{SOC hints} and \emph{no-hints}. In the \emph{SOC hints} mode we use domains from the IOC list provided by SOC as seeds. The domains output by our detector in both modes (not considering the seeds provided by SOC) were validated as follows. We first query VirusTotal and the IOC domain list to verify their status (three months after they were detected -- to allow anti-virus and blacklists to catch up). If the domain is alerted upon by at least one scanner used by VirusTotal or it's an IOC we consider it \emph{known malicious}. For other domains, we collect additional information and hand them to a security analyst for manual investigation. Specifically, we retrieve the associated URLs from the log data and crawl them to examine the responses. The URLs are also manually submitted to McAfee SiteAdvisor\ignore{\footnote{\url{http://http://www.siteadvisor.com/}}}. Based on the URLs, the response to our crawler and the result from SiteAdvisor, we classify the remaining domains into four categories: \textit{new malicious} (e.g., same URL patterns as known malicious domains, returning malicious content or flagged by SiteAdvisor), \textit{suspicious} (not resolvable when crawled, parked or having some questionable activities),  \textit{legitimate} (no suspicious behavior or code observed) and \textit{unknown} (504 HTTP response code, a sign of server error). Since we only have a few unknowns (6 in total), we remove them from the final results. When reporting our results we use several metrics of interest: TDR and FDR defined in~\secref{sec:lanlres}, and \emph{new-discovery rate} (NDR) defined as the percentage of new malicious and suspicious domains detected. Here TDR is the percentage of both known and new malicious and suspicious domains among all detected domains, and FDR = 1 - TDR.

\subsection{Results for the no-hint case}

We first evaluate the effectiveness of our detector in the \emph{no-hints} mode. We compute scores for all automated domains encountered daily. For domains whose WHOIS information can not be parsed default values for \domainage\ and \domainval\ are set at average values across all automated domains. As the first step, we vary the threshold for labeling automated connections  from 0.4 to 0.48 and present results for domains detected as \cc\ domains (with score above the threshold) in Figure~\ref{fig:emceval}(a). The results demonstrate that as we increase the threshold on automated domain scores from 0.4 to 0.48 the number of domains labeled as \cc\ drops from 114 to 19, while accuracy increases (TDR increases from 85.08\% to 94.7\%). Though FDR is higher for threshold 0.4, more malicious domains (including 23 new ones not known to VirusTotal or SOC) are detected. They can be used to seed the belief propagation stage, and therefore we fix the threshold at 0.4 to evaluate the overall effectiveness of belief propagation in this mode.

Next, we vary the threshold for domain scoring in belief propagation from 0.33 to 0.85 and the result (Figure~\ref{fig:emceval}(b)) shows that the number of all detected domains varies from 265 to 114, with TDR ranging from 76.2\% to 85.1\%. Altogether in the most challenging case (when no hint is available), we detect 202 malicious and suspicious domains in February, associated with 945 hosts. Though the majority of the detected domains are already alarmed by SOC and VirusTotal (132 for threshold 0.33), only 13 are reported in the IOC list and the remaining ones are unknown to the enterprise. More interestingly, we identified many new malicious and suspicious domains not known to the community (a total of 70 new domains for threshold 0.33 resulting in an NDR of 26.4\%). This result suggests that our detector could greatly complement existing security tools by discovering new suspicious activities. Its main advantage is that it has the ability to detect new campaigns without traces of known malicious behavior.

\ignore{Next, we change the threshold for belief propagation from 0.33 to 0.85 and the result (Figure~\ref{fig:emceval}) shows that the number of all detected domains varies from 273 to 119, with FPR ranging from 23.1\% to 14.2\%. Altogether, we detect 202 bad domains in Feburary, associated with 945 hosts. Though these 945 hosts only represent xx\% of all the hosts in AC dataset, the result is still alarming since many security products have already been deployed by the enterprise. Though the majority of the domains detected are already alarmed by SOC and VirusTotal (132 for threshold 0.33), we still identified many unknown bad domains (70 new domains for threshold 0.33 and NDR is 25.6\%). This result suggests our detector could greatly complement the existing tools with new discoveries. Within the 132 domains already alarmed by SOC and VirusTotal, only 13 are reported by SOC knowledgebase, and the remaining domains are unknown to the enterprise.}

We thoroughly examined the domains labeled as new malicious and suspicious and found several prominent and interesting clusters. Among the new malicious domains, we found 5 domains hosting URLs with the same pattern \texttt{\small /logo.gif?} later confirmed by the SOC as related to Sality worm\ignore{\footnote{\url{http://www.symantec.com/security_response/writeup.jsp?docid=2006-011714-3948-99}}}. We also found 15 domains with the same URL pattern reported by VirusTotal. \ignore{This suggests our detector helps the analyst to gain more information of specific malware. }Moreover, we identified a cluster of 10 DGA domains with none of them reported by VirusTotal and SOC, demonstrating our detector's ability in capturing new malicious campaigns. All the malicious domains are under the TLD \texttt{\small .info} and their names have 4 or 5 characters (e.g., \texttt{\small mgwg.info}). 9 out of the 10 domains hosts URLs with pattern \texttt{\small /tan2.html} and visiting them will be redirected to the remaining domain \texttt{\small 1.tv990.info}.

\ignore{We thoroughly examined the domains labeled as Malicious and Suspicious and found there exist prominent and interesting clusters. Among the Malicious domains, we found 5 domains actually host URLs with the same URL pattern \texttt{/logo.gif?}. This pattern is later confirmed by a security specialist as being related to Sality worm\footnote{\url{http://www.symantec.com/security_response/writeup.jsp?docid=2006-011714-3948-99}} We also found 15 domains detected by VirusTotal exhibit the same pattern. This suggests our detector helps the analyst to gain more information of specific malware. Moreover, we identified a cluster of 9 DGA domains with none of them reported by VirusTotal and SOC, demonstrating our detector's ability in capturing new malicious campaign. All the malicious domains are under the TLD \texttt{.info} and their names have 4 or 5 characters (e.g., \texttt{mgwg.info}). 8 out of the 9 domains hosts URLs with pattern \texttt{/tan2.html} and visiting them will be redirected to the remaining domain \texttt{1.tv990.info}.
}

We labeled legitimate a set of 63 domains belonging to categories like Ad-network, Gaming, Toolbar and Torrent Tracker. They are captured by our detector because they exhibit suspicious features, like automated connections or are registered recently. Though they do not pose serious harm to the enterprise, some of them are policy violations (e.g., Gaming, Torrent Tracker). We labeled them legitimate since we did not discover any suspicious activities, but we believe these domains still need to be vetted.

\subsection{Results for the SOC hints case}

We also present results in Figure~\ref{fig:emceval}(c) for the belief propagation algorithm in \emph{SOC hints} mode seeded with 28 IOC domains. We do not include the seed domains in the results. We set the domain score threshold at 0.4 for automated domains and vary the similarity score between 0.33 and 0.45. In total, we detect between 137 domains (at threshold 0.33) to 73 domains (at threshold 0.45), with TDR ranging from 78.8\% to 94.6\%. Among the 137 detected domains, 108 turn out to be malicious (either known or new) and suspicious, which is about four times larger than the set used for seeding.

\ignore{for automated connection threshold of 0.4 and domain score in belief propagation varying from 0.33 to 0.45. We used 28 domains detected by SOC during the month of February as seeds, which we don't include in the results. In total, we detect between 137 domains (at threshold 0.33) to 73 domains (at threshold 0.45), with TDR ranging from 78.8\% to 94.6\%. Among the 137 detected domains, 108 turn out to be malicious (either known or new) and suspicious, which is about four times larger than the set used for seeding.
}

%The overall accuracy turns out to be higher than in the no-hint setting, since false positives are less likely to be included in SOC hints.

%which is about four time larger as much as the domains used as seeds.

%interesting cases

Among the 108 malicious and suspicious domains, 79 domains are confirmed by SOC or VirusTotal, leaving 29 domains as our new findings. We inspect the new findings and identify an interesting group of domains generated through DGA. This group consists of 10 domains under TLD \texttt{\small .info} and the name for each domain has 20 characters (e.g.,  \texttt{\small f0371288e0a20a541328.info}). Surprisingly, the registration dates for most of the domains are later than the time when we detected them. For example, one domain is detected on 2014/02/13 but registered on 2014/02/18. Attackers use DGA domains to increase the robustness of their \cc\ centers against take-downs, and they only register a portion of the domains to reduce the cost. Our detector is able to detect the malicious domains before registeration and obtain an advantage in the arm-race.

\ignore{
Among the 108 bad domains, 79 domains are confirmed by SOC or VirusTotal, leaving 29 domains as our new findings. We insepct the new findings and identify an interesting group of domains generated through DGA. This group consists of 10 domains under TLD \texttt{\small .info} and the name for each domain has 20 characters (e.g., f0371288e0a20a541328.info). Surprisingly, the registration dates for most of the domains are later than the time when detected by us. For example, one domain is detected on 2014/02/13 but registered on 2014/02/18. Attackers use DGA domains to increase the robustness for malware under taken-down efforts from security industry, and they only register a portion of the domains to reduce the cost. Our detector is able to detect these ``ghost'' domains and take the advantage in the arm-race.}

Finally, we compare the results of the two modes of operation. Only 21 domains are detected in both modes, which is a small portion compared to 202 and 108 malicious and suspicious domains detected separately. When deployed by the enterprise, we suggest our detector configured to run in both modes, in order to have better coverage. We present two case studies for both modes of operation in the Appendix. As we have shown, starting from a seed of known malicious domains or hosts, the algorithm in  \emph{SOC hints} mode can identify suspicious domains with high accuracy. The \emph{no-hint} case has the unique capability of identifying new unknown attack campaigns, especially \cc\ communications of these campaigns (even when only a single host is compromised). We recommend that \cc\ detected domains are first vetted by the SOC and then the algorithm can be used in the \emph{SOC hints} mode for those confirmed malicious domains.

Both variants include configurable thresholds for scoring automated and non-automated domains. These thresholds can be chosen by the SOC according to the capacity of the team performing manual investigation, and various tradeoffs between accuracy and larger coverage.

\section{Related Work}

%\vspace{2pt}

%\noindent {\bf Detection of \cc\ communication.} Our work focuses on the initial phases of APT attacks and the detector we built is capable of capturing malicious domains used by attackers, including domains for \cc\ communication. \cc\ domains are important components of botnet infrastructure and are studied widely. A number of studies attempt to find spatial-temporal correlations between hosts independent of the \cc\ communication  protocol~\cite{BotSniffer,BotMiner,BotFinder, Yen08}. DISCLOSURE~\cite{Bilge2012} presents a set of detection features to identify \cc\ traffic using NetFlow records, while Paxson et al.~\cite{Paxson2013} detect malicious communication established through DNS tunnels.

%Both DISCLOSURE and BotFinder identify regular (periodic) communication in C&C channels, but we introduce a novel and efficient method resilient to outliers.

%A different approach is to run malware samples in a sandbox environment and extract characteristics of the \cc\ communication. Jackstraws~\cite{JACKSTRAWS} extracts behavior graphs of system calls for malware processes and compares new binaries with known malware templates. ExecScent~\cite{ExecScent} generates templates of control protocols from a feed of malware inteligence data, and subsequently uses them to detect new suspicious activities in enterprise network. Our approach is different in not relying on existing malware samples or traces to bootstrap the detection and is therefore able to capture \cc\ domains used by new malware campaigns.

Our work focuses on detecting early-stage infections within enterprise perimeters, including communications related to malware delivery and \cc. There has been a large body of work in this area, but to the best of our knowledge, we are the first to exploit the relationship between malicious domains associated with the same attack campaign, and to detect them by a graph-theoretic framework based on belief propagation. We describe here related work in the literature.

\noindent {\bf Detection of \cc\ communication.}  \ignore{\cc\ domains are important components of botnet infrastructure and there is an extensive body of research on detecting them.} Some of the previous work detecting \cc\ domains in botnets require malware samples as input to detect connections with similar patterns (e.g., BotFinder\cite{BotFinder}, Jackstraws\cite{JACKSTRAWS}). Anomaly-based botnet detection systems (e.g., BotMiner\cite{BotMiner}, BotSniffer\cite{BotSniffer} and TAMD~\cite{Yen08}) typically detect clusters of multiple synchronized hosts infected by the same malware. In contrast to these, our approach does not require malware samples and can detect a single compromised host contacting the \cc\ server.

DISCLOSURE~\cite{Bilge2012} identifies \cc\ traffic using features extracted from NetFlow records but incorporates external intelligence sources to reduce false positives. Our \cc\ detector is different in that it leverages enterprise-specific features extracted from HTTP connections. From that perspective, ExecScent~\cite{ExecScent} is close to our work in detecting \cc\ communications in large enterprise network. However, ExecScent needs malware samples to extract templates representing malicious \cc\ connections. The templates are adapted to a specific enterprise considering the popularity of different features (URL patterns, user-agent strings, etc.). Our work complements ExecScent in detecting new unknown malware that can be provided as input to the template generation module.

%The templates are composed of patterns summarized from URL, User Agent and are only capable of capturing communications from known malware. Our work could complement ExecScent in detecting unknown and new malware.

\vspace{2pt}

\noindent {\bf Detection of malware delivery.} \ignore{Malware delivery is another important stage within the infection process addressed in previous work. } Nazca~\cite{invernizzi2013miab} analyzes web requests from ISP networks to identify traffic related to malware delivery and unveils malicious distribution networks. CAMP~\cite{Rajab2013} determines reputation of binary downloads in the browser and predicts malicious activities.
BotHunter~\cite{BotHunter} identifies sequences of events during infection, as observed from a network perimeter. Our approach does not depend on specific events occurring during infection (which can be changed easily or may not be observable), but more focused on detecting related malicious domains and compromised hosts.

%BotHunter~\cite{BotHunter} identifies sequences of events that take place during infection, as observed from a network perimeter. This includes leveraging IDS logs to detect inbound scans or exploits, followed by malware delivery, \cc\ communication, and outbound scans.  Our work looks for different stages of host compromise (e.g., delivery, \cc) and leverages the relationship between multiple malicious domains contacted by a host during infection.

\ignore{Malware delivery is another important stage within the infection process and several research works have been proposed. Invernizzi et al. proposed Nazca~\cite{invernizzi2013miab} which looks into the collective traffic related to binary downloading and then identifies the malware distribution network. Rajab et al. proposed CAMP~\cite{Rajab2013} utilizing reputation-based detection to identify malware. Our work oversees different stages in infections (both \cc\ and delivery). By leveraging the relationships between domains, we are able to expand the detection result across stages.}

\noindent {\bf Detection of malicious domains.}
Domains used in malicious activities are backed by highly resilient infrastructures to deal with takedowns or blacklisting, and hence exhibit unique characteristics distinct from benign sites. Another branch of work detects domains involved in malicious activities \ignore{by these unique connection pattern, such as those } by patterns observed in DNS traffic (e.g., EXPOSURE~\cite{Bilge2011}, Notos~\cite{Antonakakis2010}, Kopis~\cite{Antonakakis2011}, and Antonakakis et al.~\cite{Antonakakis2012}). Paxson et al.~\cite{Paxson2013} detect malicious communication established through DNS tunnels. Carter et al.~\cite{Carter14} use community detection for identifying highly localized malicious domains in the IP space. \ignore{Ma et al.~\cite{MaKDD09} identifies suspicious URLs by examining lexical properties of the URL and information about the domain, such as WHOIS information, reputation, geographic location, and TTL value in DNS responses.}

\ignore{
Another branch of related literature detects malicious domains not necessarily related to infections. For example, EXPOSURE~\cite{Bilge2011}, Notos~\cite{Antonakakis2010}, and Kopis~\cite{Antonakakis2011} perform passive DNS analysis to identify malicious domains, while \cite{Antonakakis2012} detects DGA domains. Paxson et al.~\cite{Paxson2013} detect malicious communication established through DNS tunnels. Carter et al.~\cite{Carter14} use community detection for identifying highly localized malicious domains in the IP space.
}

\vspace{2pt}
\noindent {\bf Anomaly detection in enterprise network.} Beehive~\cite{Beehive} is an unsupervised system identifying general anomalies in an enterprise setting including policy violations and malware distribution. Our work is specifically targeting enterprise infections which pose high risk and potential financial loss.

\vspace{2pt}

%\noindent {\bf Targeted attacks.} Widely-used techniques for detecting malicious activities (e.g., intrusion-detection systems~\cite{Snort}) are quite ineffective against well-orchestrated APT campaigns, as APT attackers customize their tools to specifically evade existing defenses. The recent Verizon report~\cite{DIBR14} explicitly states that 85\% of targeted attacks are discovered by external means, and not by security products.

%Starting with Stuxnet~\cite{Stuxnet} APTs have become an emergent threat posing risks to intellectual property, critical infrastructure and national security. Hutchins et al.~\cite{LMAPT} introduce the first model of APT lifecycle (or kill chain), describing the typical stages of an APT attack\ignore{ and advocates for an intelligence-driven defensive model}. Information about attackers' tools, techniques and tactics has been revealed in a report published by MANDIANT which focuses on one of the most prominent APT groups (dubbed ``APT1''). Thonnard et al.~\cite{Thonnard12} studied the characteristics of 26,000 social-engineering targeted attacks in 2011, and showed that at least eight of them exploited zero-day vulnerabilities. We are aware of only several papers describing ideas on detecting APT campaigns~\cite{NVictims,Giura13}, but they lack rigorous evaluation.

\noindent {\bf Targeted attacks.}  The threats in cyberspace keep evolving and more sophisticated attacks recently emerged. Some targeted attacks (APT) are well-funded, carefully orchestrated and persist in the victim environments for years before detection.

%Some famous campaigns like those initiated by the APT1 group~\cite{APT1} aim to steal high-profile information from enterprises and government organization, causing high damage.

Detecting targeted attacks in general is a very challenging task. These attacks are usually very stealthy and able to evade existing defenses~\cite{DIBR14}. However during the automated infection stage many campaigns (e.g., Shady RAT~\cite{Shady}, Mirage~\cite{Mirage}, APT1~\cite{APT1}) exhibit similar infection patterns. Recent studies have shown that even though in theory APTs could be arbitrarily sophisticated, in practice goal-oriented attackers use relatively low levels of sophistication~\cite{Thonnard12, Marczak2014, LeBlond2014}. We leverage some common patterns observed during the infection stage to build a detector tailored to an enterprise. Our detection result on the LANL's APT infection discovery challenge indicates that our techniques have potential in detecting infections originated from targeted attacks.

\section{Limitations and Discussion}

Attackers could in principle use a number of techniques to evade our detectors. For instance, they may communicate through different channels than HTTP/HTTPs  but other ports are commonly blocked at enterprise borders. Attackers could also compromise popular web sites for delivering malicious payloads or \cc\ communications. However, this is not broadly observed. The analysis of traffic to popular sites require different models and assumptions, due to the amount of (benign) noise. \ignore{Since our detector is not monitoring the traffic between hosts, we may miss communications between hosts during various attack stages. This requires other types of data for detection, e.g., internal DNS or NetFlow data. }\ignore{This again is less likely to happen as attackers have to evade monitors and firewalls deployed inside the network. }As another evasion technique, attackers can randomize timing patterns to \cc\ servers, but according to published reports (\cite{APT1}) this is uncommon. Our dynamic histogram method is resilient against small amounts of randomization introduced by attackers. Detecting \cc\ communication with completely randomized timing patterns (without having access to malware samples and without correlating activity from multiple hosts) is a challenging problem to the community. Nevertheless, we believe that the infection patterns that we detect are quite prevalent in many attacks.

\ignore{
some features are quite resilient to changes in attacker behavior. For instance, we believe that hosts compromised by attackers in the same campaign will exhibit certain similarity in connections to rare domains. In principle  attackers can use a different set of domains for communicating with every individual host, but this would largely affect the attacker's operating cost. The sequence of connections to these domains is also a resilient indicator as the host needs to download the payload first and then communicate with the \cc\ center (following the APT kill chain). It is also likely that attackers use recently registered domains (unless they compromise legitimate ones), and a small number of IP subnets. Timing patterns across different hosts can be randomized relatively easily by attackers. Our dynamic histogram method is resilient against small amounts of randomization, but it is still tailored at detecting almost periodic \cc\ traffic.  Nevertheless, the method is extensible to a variant in which we can detect if two hosts employ similar inter-connection distributions to the same \cc\ domain (for an arbitrary distribution like normal, exponential, gamma, etc.). We leave this extension for exploration in future work.
}

The approach we proposed is meant to complement the existing tools rather than replace them. The results from~\secref{sec:eval_emc} demonstrate that our belief propagation algorithm in both variants (\emph{SOC hints} and \emph{no-hint}) detects new suspicious activities overlooked by deployed defense mechanisms. These include both domains associated with existing malware campaigns (and identified by VirusTotal), but with new presence in the enterprise of our study, as well as entirely new malware campaigns (not yet detected by anti-virus technologies). Since our methods are focused on detecting the initial infection stages of a campaign it is difficult to determine how many of these suspicious activities are related to more advanced attacks, and how many are mainstream malware variants. We believe that monitoring activity to these suspicious domains over longer periods of time, as well as correlating with information from other data sources will answer this question, and we leave this as an interesting avenue for future work.

\bibliographystyle{abbrv}
% size options, uncomment one
%\footnotesize
{\scriptsize
\bibliography{refs}}
%\balancecolumsn

\appendix

\section{Case studies}
\label{sec:case}

Below we report two interesting communities of malicious domains and compromised hosts detected by our techniques.

\begin{figure}[thbp]
%\vspace{-0.8cm}
\begin{center}
  \begin{tabular}{lr}
  \includegraphics[width=3.3in]{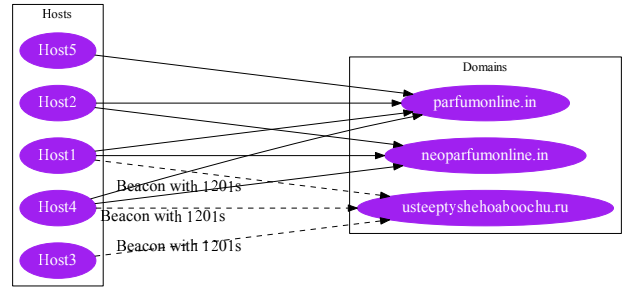}
  \end{tabular}
\end{center}
%\vspace{-0.8cm}
\caption{Example community of compromised hosts and detected domains in \emph{no-hint} mode on 2/13.}
%todo:work on title later
%\vspace{-.5cm}
\label{fig:case4}
\end{figure}

Figure~\ref{fig:case4} illustrates an example of a community of malicious domains detected in the \emph{no-hint} mode on 2/13. We start from detecting suspicious \cc\ domains (24 domains are identified that day). We pick for illustration one (\texttt{\small usteeptyshehoaboochu.ru}) contacted by three hosts regularly at the same period (120 seconds). This domain is confirmed by VirusTotal and is associated with a malware labeled by Sophos as \texttt{\small Troj/Agent-AGLT}.  \ignore{We inspected the URLs used by hosts when contacting this domain and found an interesting pattern of the form \url{gate.php?user=*&id=*&type=*&hashrate=*}, with unique parameter values for each host.} Starting from this \cc\ domain, belief propagation iteratively discovers two other domains confirmed by VirusTotal (likely the delivery stage) and two other hosts connecting to them.

\ignore{
We inspect the URLs under this domain and feel it could be used to keep track of the compromised host: the URLs are formed like \url{gate.php?user=xxx&id=xxx&type=xxx&hashrate=xxx}, while the paramter values are unique to each host. Iteratively, starting from these 3 hosts, we discover 2 other domains (both of them confirmed by VirusTotal) and 2 other hosts connecting to them.
}

%2014-02-13 12:20:17	usteeptyshehoaboochu.ru	http://usteeptyshehoaboochu.ru/gate.php?user=dc77443fa8124caaa17b2923433bb082&id=1001&type=3&hashrate=7.72	None	None

\begin{figure}[thbp]
\begin{center}
  \begin{tabular}{lr}
  \includegraphics[width=3.3in]{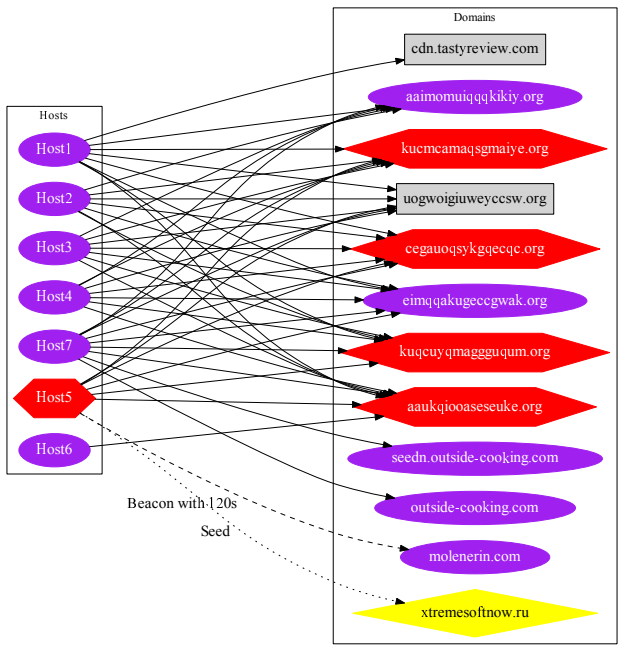}
  \end{tabular}
\end{center}
%\vspace{-0.8cm}
\caption{Example community of compromised hosts and detected domains in \emph{SOC hints} mode on 2/10. The yellow diamond-shape domain is used as seed, purple ellipse-shape domains are detected by VirusTotal and red hexagon-shape domains are confirmed malicious by SOC. The grey rectangle-shape domains are not confirmed by any existing tools at the time of detection. The hostnames are anonymized. Red hexagon-shape hosts are confirmed by SOC and purple ellipse-shape ones are other compromised hosts.}
%todo lz:haven't figured out how to add legend into dot graph, work on it later
%\vspace{-.5cm}
\label{fig:case2}
\end{figure}

Figure~\ref{fig:case2} illustrates a community of domains detected in the \emph{SOC hints} mode on 2/10. To bootstrap the detection, we use domain \texttt{\small xtremesoftnow.ru} from the SOC database as a seed. This appears to be a \cc\ server for Zeus botnet\ignore{~\footnote{\url{https://zeustracker.abuse.ch/monitor.php?host=xtremesoftnow.ru}}}. The domain is accessed by \texttt{\small Host 5} which is confirmed infected by the SOC team. \texttt{\small Host 5} contacted 7 domains registered under TLD \texttt{.org}. Four of them are confirmed by SOC and also reported by Sophos as contacted by different  malwares (\texttt{\small Troj~Ramdo-B}, \texttt{\small Troj~Ramdo-K}, \texttt{\small Troj~Ramdo-V} and \texttt{\small Troj~Symmi-S}). This indicates that the compromised machine downloaded additional malware after the initial compromise. Two of the remaining \texttt{.org} domains are not confirmed by SOC but alarmed by VirusTotal. Interestingly, one domain ({\tt \small uogwoigiuweyccsw.org}) has a similar naming pattern with other malicious domains, but is not picked up by either SOC or VirusTotal. This is an example of new discovered domain. The second iteration of belief propagation discovers six additional hosts contacting a similar set of domains as \texttt{\small Host 5}, indicating that they could be infected with the same malware. Besides, four other suspicious domains are also identified as contacted by these hosts, with three of them confirmed by VirusTotal (including one automated domain) and only one (\texttt{\small cdn.tastyreview.com}) being legitimate.

%These examples suggest that our techniques can help security analysts identify the root cause of security incidents, as well as automatically infer the domains and hosts associated with an entire attack campaign.

\ignore{This indicates attackers instructed the compromised machine to excessively download malwares after making it a bot. Two of the remaining \texttt{org} are not confirmed by SOC team but alarmed by VirusTotal and there is even one domain not alarmed by any tools when we run our detector. Clearly, the name of this domain follow the similar pattern as other \texttt{org} domains and is likely to be registered using the same domain generation algorithm (DGA). Therefore we also mark it a true positive. In the next iteration, 6 other compromised hosts are discoverd as they connected to the similar set of domains as \texttt{host5}, indicating they could be infected with the same malware. This suggests our tool is very helpful for the security analysts to identify the root cause of security incidents. Besides, 4 other domains are also identified as connected by these hosts, with 3 of them confirmed by VirusTotal (including one beaconing domain) and only 1 domain to be false-positive.}

%\section{Headings in Appendices}

% That's all folks!
\end{document}